\documentclass[11pt,a4paper]{article}

\usepackage[T1]{fontenc}
\usepackage{amsmath}
\usepackage{amsthm}
\usepackage{amsfonts}
\usepackage{amssymb}
\usepackage{graphicx,subfigure,color}
\usepackage{hyperref}
\usepackage[font=footnotesize,labelfont=bf]{caption}
\usepackage{amscd}
\usepackage{enumerate}
  \usepackage{cite}

\def\be{\begin{equation}}
\def\ee{\end{equation}}
\def\bea{\begin{eqnarray}}
\def\eea{\end{eqnarray}}

\begin{document}
\def\thefootnote{\fnsymbol{footnote}}

\begin{center}
\Large{\textbf{Inflation and late time acceleration designed by Stueckelberg massive photon}} \\[0.3cm]

\large{\"{O}zg\"{u}r Akarsu$^{\rm a}$, Metin Ar{\i}k$^{\rm b}$, Nihan Kat{\i}rc{\i}$^{\rm a}$}
\\[0.3cm]

\small{
\textit{$^{\rm a}$ Department of Physics, {\.I}stanbul Technical University, 34469 Maslak,  {\.I}stanbul, Turkey}}

\small{
\textit{$^{\rm b}$ Department of Physics, Bo\u{g}azi\c{c}i University, 34342 Bebek, {\.I}stanbul, Turkey}}


\end{center}

\vspace{.6cm}

\hrule \vspace{0.2cm}
\noindent \small{\textbf{Abstract}\\
We present a mini review of the Stueckelberg mechanism, which was proposed to make the abelian gauge theories massive as an alternative to Higgs mechanism, within the framework of Minkowski as well as curved spacetimes. The higher the scale the tighter the bounds on the photon mass, which might be gained via the Stueckelberg mechanism, may be signalling that even an extremely small mass of the photon which cannot be measured directly could have far reaching effects in cosmology. We present a cosmological model where Stueckelberg fields, which consist of both scalar and vector fields, are non-minimally coupled to gravity and the universe could go through a decelerating expansion phase sandwiched by two different accelerated expansion phases. We discuss also the possible anisotropic extensions of the model. } 
\\
\noindent
\hrule
\def\thefootnote{\arabic{footnote}}
\setcounter{footnote}{0}

\let\thefootnote\relax\footnote{\textbf{E-Mail:} akarsuo@itu.edu.tr, metin.arik@boun.edu.tr, nihan.katirci@itu.edu.tr}

\def\thefootnote{\arabic{footnote}}
\setcounter{footnote}{0}

\section{Introducing Stueckelberg mechanism}
\label{intro}
The observation of only left handed weak interactions implies the violation of parity symmetry in nature. Thus fermions as well as gauge bosons should have been massless to preserve gauge symmetry. This was of course in gross contradiction with the experiments demonstrating that the weakly interacting fermions and gauge bosons are massive, though the neutrinos first were assumed to be massless but later it was shown that they are also massive. Indeed gauge bosons mediating the weak interaction were naturally expected to be massive since weak force is a short range force. The clever solution to this issue was to introduce a Higgs field which transforms as a doublet under the weak $\rm SU(2)_L$ symmetry so that $W^+$, $W^-$ and $Z^0$ bosons as well as all the charged fermions would be massive \cite{Weinberg:1979pi}. Indeed, it has later been realised that neutrino is the lightest particle in the Standard Model (SM) with a mass smaller by at least three orders of magnitude than the electron mass. The 2015 Nobel Prize in Physics was given to the discovery of neutrino oscillations that shows neutrinos are massive. Therefore the SM should has been modified in order to give a natural explanation to the question why neutrino masses are so small but non-zero. A similar modification that makes neutrinos massive may be valid for photon. As dictated by Okun, ``\emph{such a small photon mass, albeit gauge non-invariant, does not destroy the renormalizability of Quantum Electrodynamics (QED) \cite{Feldman:1963zz,itzykson80} and its presence would not spoil the agreement between QED and experiment. This also motivates incessant searches for a non-vanishing tiny photon mass"} \cite{Okun:1991nr,Okun:2006pn}. Historically there have been proposals on photon being massive by renowned physicists such as Einstein \cite{Einstein:1905cc,Einstein:1917zz}, de Broglie \cite{deBroglie:1922zz,debroglie23,debroglie34,debroglie40,DeBroglie:1972hj}, and Schr\" odinger \cite{schro55}. The possibility of photon having an ultralight mass is present in de Broglie's doctoral thesis (1924) \cite{debroglie-thesis}, which was preceded by the two published articles of him; he first time mentioned a massive photon in 1922 \cite{deBroglie:1922zz} and provided an estimate on its mass in 1923 \cite{debroglie23}. However this idea is often ascribed to one of his doctorate students Proca who introduced the so-called Proca equation for a massive vector field in Ref. \cite{Proca:1900nv} in relation with series of his articles published in 30s \cite{proca36a,proca36b,Proca:1900nv,proca36d,proca37}. We note, however, that the main aim of Proca in these works was the description of spin-1/2 particles inspired by the neutrino theory of light due to de Broglie \cite{proca88,borne01}.

In classical electromagnetism, encoded in the Maxwell theory, the Lagrangian density for the canonical electromagnetic field is given as follows:
 \bea & {\cal L}_{\rm EM} =& - \frac14
F_{\mu\nu} F^{\mu\nu}, \label{EM} 
\eea
where $F^{\mu\nu}$ is the electromagnetic field strength tensor and there is no mass term. It should be noted here that we use natural units with $\hbar=c=1$ throughout this mini review. It had been believed by many that only massless vector field theories are gauge invariant. The massive spin-1 vector field is given by Proca Lagrangian density
 \bea & {\cal L}_{\rm Proca} =& - \frac14
F_{\mu\nu}(V) F^{\mu\nu}(V) + \frac12 m^2 V^\mu V_\mu, \label{proca} 
\eea
where $V$ is the Proca field and $m$ is its mass that breaks the gauge symmetry. Lorentz condition for the Proca field $\partial^{\mu} V_{\mu}=0$ is automatically satisfied and the Hamiltonian density is positive definite. In commutation relations $1/m^2$ term leads to quadratic ultraviolet (UV) divergence, which can not be eliminated by renormalisation. Stueckelberg, on the other hand, achieved Lorentz covariance and even gauge invariance by introducing an auxiliary scalar field that contributes to the longitudinal polarisation of the massive photon of Proca theory in addition to the two transverse polarisations, known as helicities (original paper \cite{stueckelberg:1957}, see \cite{Ruegg:2003ps} for a detailed review). The so called Stueckelberg Lagrangian density is given as follows;
\bea & {\cal L}_{\rm Stu} =& - \frac14
F_{\mu\nu} F^{\mu\nu} + \frac12 m^2 \left( A_\mu- \frac1m
\partial_\mu B\right)^2, \label{stu} \eea
which is invariant under gauge transformation given by
\be
A_{\mu} \rightarrow A_{\mu}+\partial_{\mu}\lambda\quad \textnormal{and}\quad B \rightarrow B+m\lambda
\ee
for arbitrary $\lambda$. Stueckelberg fields consist of a gauge vector field $A_{\mu}$ and the Stueckelberg scalar $B$, and these two fields share the same mass $m$. In Stueckelberg mechanism Lorentz condition is not automatically satisfied just like in standard QED, but spurious scalar field is introduced so as to contribute to the commutation relation and make the Hamiltonian density positive definite. While the derivatives in the commutator make the theory singular at higher energies, the absence of the term $1/m^2$ and of the derivative terms in the commutation relations of the vector fields in Stueckelberg mechanism save the renormalisability. It is worth to mention here also that massive abelian gauge theories with Stueckelberg mechanism are proved to be renormalisable and unitary \cite{Lowenstein:1972pr} (see also \cite{vanHees:2003dk}).

We can safely say that there is no need for the Higgs mechanism (spontaneous symmetry breaking) to give mass to the abelian gauge fields, the Stueckelberg mechanism does so. Stueckelberg model can be considered as the free abelian Higgs model \cite{Kibble65},
\bea
&{\cal L}_{\rm FH} =& -\frac14 F_{\mu\nu}F^{\mu\nu} +\left |(\partial_\mu -i e A_\mu) \Phi
\right|^2,
\eea
where the magnitude of the complex scalar field is fixed with the mass, and its phase corresponds to the Stueckelberg scalar field given by 
 \bea
 \Phi = \frac1{\sqrt2} \frac{m}{e}\,{\rm e}^{\frac{ieB}{m}},
 \eea
where $e$ is the electron charge. This would also provide an extension of the SM but unfortunately this trick can be used as a mass mechanism only for abelian vector field theories. Although there have been many attempts to make the non-abelian vector fields massive via Stueckelberg mechanism, this could have been never succeeded and Higgs mechanism is still the only way for making non-abelian vector fields massive. To quantise the Stueckelberg fields, gauge fixing term should be added and then the Lagrangian density can be written as
\bea
{\cal L}_{\rm Stu} = -\frac14
F_{\mu\nu} F^{\mu\nu} + \frac12 m^2 \left( A_\mu- \frac1m
\partial_\mu B\right)^2 - \frac{1}{2\alpha} \left( \partial^\mu A_\mu+ m B\right)^2, \label{stu} \eea
and the invariance of the gauge fixing term requires that
\be
(\partial^2+\alpha m^2)\lambda=0,
\ee
where $\alpha$ is the gauge fixing parameter. Stueckelberg Lagrangian density is in fact the Proca Lagrangian density supplemented by the 't Hooft gauge fixing term in the Stueckelberg-Feynman gauge $\alpha=1$ \cite{taylor76}. Proca and Stueckelberg theories are related by \cite{Pauli:1941zz},
 \be 
 \label{procastu}
{\cal L}_{\rm Stu}+\frac12 \left( \partial^\mu A_\mu+ m B\right)^2={\cal L}_{\rm Proca}.
\ee
through the following transformation between fields $V_{\mu}$ and $A_{\mu}$
 \be
 V_\mu
\equiv A_\mu -\frac1m \partial_\mu B. \label{220}
\ee 
Stueckelberg mechanism has had various applications in theoretical physics that go far beyond its original motivation, e.g., in gravity-included fundamental theories such as superstring theories. Delbourgo supersymmetrized $(N=1)$ the Stueckelberg mechanism long time ago \cite{Delbourgo:1975uf}. Guerdane et al. extended this mechanism to supersymmetric theories \cite{Guerdane:1991at}. However, the required higher derivatives in the gauge fixing terms to generate equal numbers of bosonic and fermionic degrees of freedom can lead to ghosts in these types of theories. Deser and van Nieuwenhuizen \cite{Deser:1974cz} then argued that the higher order theories would be renormalizable \cite{Stelle:1976gc} but contain ghosts. Sezgin and van Nieuwenhuizen \cite{Sezgin:1981xs} then showed that ghost free higher derivative gravity theories are also possible. Bergshoeff and Kallosh \cite{Kallosh:1990vq} quantized the superparticle in 10-dimensional superstring theory using BRST invariance. The Stueckelberg transformation found by Fisch and Henneaux \cite{Fisch:1989rn} simplifies this quantization procedure. Using this Stueckelberg symmetry, Bergshoeff and Kallosh then could write BRST invariant free quadratic Lagrangian without any constraints \cite{Bergshoeff:1990vg}. Delbourgo and Salam \cite{Delbourgo:1975aj} and long time after Arkani-Hamed \cite{ArkaniHamed:2002sp} applied the Stueckelberg mechanism to the graviton field. The Stueckelberg mechanism is widely used in the formulation of the antisymmetric partner to the graviton \cite{Kalb:1974yc} and in covariant string field theory \cite{Marshall:1974wf}. It is quite interesting that longitudinal components of massive graviton (that gained mass via Stueckelberg trick) couple to the matter distribution and lead to the van Dam Veltman Zakharov (vDVZ) discontinuity \cite{vanDam:1970vg,Zakharov:1970cc,Boulware:1973my}: Contrary to popular belief, predictions on the gravitational bending of light trajectories are the same in massive gravity and in the standard General Relativity (GR), however Newtonian potentials of massive gravity are not consistent with that of GR and that of zero mass limit of the massive gravity. Porrati showed that this vDVZ discontinuity is automatically removed from Pauli-Fierz theory in curved spacetime \cite{Kogan:2000uy,Porrati:2000cp} (for a detailed review about massive gravity theories, see \cite{Hinterbichler:2011tt}). 

Recently Belokogne and Folacci \cite{Belokogne:2015etf} presented a detailed discussion on the Stueckelberg massive electromagnetism on an arbitrary four-dimensional curved spacetime, which permits, contrary to Proca type massive electromagnetism, to the construction of the expectation value of the renormalised stress-energy tensor associated with the Stueckelberg theory. Doing so, one is allowed to write this expectation value that appears as a source in the semiclassical Einstein equations which in return govern the backreaction of the quantum field theory on the spacetime geometry. On the other hand, backreaction has been worked in different studies within the framework of Robertson-Walker spacetime. For instance for massless/light non-minimally coupled scalar fields, backreaction from inflationary quantum fluctuations was studied in references \cite{Glavan:2013mra,Glavan:2015cut} and in these papers, authors also discuss the late time predictions on the backreaction of the model. They point out the need for an extension of this analysis using different fields such as gauge fields. In \cite{Prokopec:2002jn,Prokopec:2002uw,Prokopec:2003bx} photons are produced in a locally de-Sitter space via a light, minimally coupled, charged scalar field, in contrast to the Higgs mechanism, where the Proca mass is induced by scalar condensate. We know that the original Stueckelberg model includes the coupling of gauge fields to a light scalar field and this seems interesting enough to motivate one to investigate the backreaction from gauge fields. These fields may even be non-minimally coupled to gravity as we consider in this mini review.

Although the application of the Stueckelberg mechanism to gravity included field theories are quite common, there are only few applications of it in cosmology \cite{Chimento:1990dk,Frob:2013qsa,Akarsu:2014eaa,Kouwn:2015cdw,Belokogne:2016dvd}. It is obvious that if the electromagnetic field is massless then it is the only long range interaction which could be relevant on cosmological scales apart from gravity. On the other hand, even if it is massive, its mass can be sufficiently small to allow us to consider it as a long range force, namely, a force that could be effective at cosmological scales. It should be noted here also that if the mass of the Stueckelberg fields is very tiny, then its possible cosmological consequences may be the only opportunity for testing this mechanism. A motivation behind the cosmological construction of a relation between the sufficiently tiny photon mass and the late time acceleration is that the lowest value of possible photon mass that can be directly measured is approximately in the same order with the corresponding mass scale of the Hubble constant, which is $\sim10^{-33}$ eV. Although there is no evidence for a non-zero photon mass, it is possible that massive theory yields so small mass so that it cannot be distinguished from massless theory. But concept of massive photon is not new as we discussed above, and it can appear in fundamental unified theories, e.g., it was predicted by string theory in the 90s \cite{Kostelecky:1991ak,Kostelecky:1990pe}.

We know that gauge transformation does not change the physics and hence experimental bounds on the Proca type mass should be valid for the bounds on the Stueckelberg type mass. Heisenberg's uncertainty principle says that it is impossible to do any experiment that would firmly establish that the photon rest mass $m$ is exactly zero but the lowest value for any mass dictated by the uncertainty principle, namely, $m \geq 1/ {\rm \Delta} t \sim 10^{-33}$ eV ($10^{-66}$ g) for the age of the present universe $ {\rm \Delta} t \sim 10^{10}$ yr. Hence it is legitimate to search for the lowest upper bounds on photon mass above this value observationally either by short range laboratory or long range astrophysical and cosmological measurements (for a list of limits see \cite{Goldhaber:2008xy}). From laboratory tests, it is found in \cite{Williams:1971ms} for instance that the measurements of deviations from the Coulomb law potential considering a modified potential of the form $1/r^{2+q}$ gives $q= (2.7\pm3.1) \times 10^{-16}$ corresponding to a photon mass limit $ m \lesssim 10^{-14}$ eV. Such laboratory upper limits are usually several orders of magnitude larger than those of astrophysical observations. The upper bound on $m$ is about $10^{-15}$ eV from the measurements of Earth's magnetic field \cite{Fischbach:1994ir} and Pioneer-10 measurements of Jupiter's magnetic field \cite{Davis:1975mn}. Using the magnetohydrodynamics argument concerning survival of the Sun's field to the radius of the Earth's orbit (1 AU), the upper bound on the photon mass was given as $m \lesssim 5 \times 10^{-17} $ eV in \cite{Ryutov:1997zz} and this was accepted by the Particle Data Group (PDG) \cite{Eidelman:2004wy} in 2004.  The extension of this study to the distance that of the Pluto's orbit (40 AU) improved the limit almost two orders of magnitude as $m \lesssim 8 \times 10^{-19} $ eV in 2007 \cite{Ryutov:2007zz} and also set, in 2008 \cite{Amsler:2008zzb}, the currently accepted limit by the PDG  \cite{Olive:2016xmw}. On the other hand, it has been recently shown using Cluster four spacecraft data in the solar wind at 1 AU that the upper bound for $m$ lies between $8\times 10^{-14}$ eV and $2\times 10^{-15}$ eV \cite{Retino:2013gga}, and using the frequency-dependent time delays in fast radio bursts that it is $1.8\times 10^{-14}$ eV \cite{Bonetti:2016cpo}, which make the estimates \cite{Ryutov:1997zz,Ryutov:2007zz} accepted by the PDG \cite{Amsler:2008zzb,Olive:2016xmw} controversial. There are also much more stringent limits such as $m \lesssim 10^{-27}$ eV \cite{Chibisov:1976mm,Lakes:1998mi} and $m \lesssim 10^{-26}$ eV \cite{Adelberger:2003qx} from the galactic magnetic field if the galactic magnetic field is in the Proca regime, though these might be considered speculative (see \cite{Goldhaber:2008xy,Goldhaber:1971mr} for details). If photon has a rest mass, either in Proca theory or in Stueckelberg theory, low frequency photons would arrive later than high frequency ones to the observer, and hence also the farther its source, the greater dispersion is measured, since the group velocity reads as follows:
\be
v_g=\sqrt{1-\frac{m^2}{\omega^2}},
\ee
where the $\omega$ is the frequency of the photon. Thus, instead of looking at high frequency photons, measurements of short time structures in lower frequency radio emissions from sources at cosmological distances would be more useful for searching photon mass. For instance fast radio bursts \cite{Bonetti:2016cpo,Wu:2016brq} are recently preferred compared with gamma-ray bursts and active galactic nuclei. It is recently proposed in \cite{Bonetti:2016cpo,Bentum:2016ekl} to investigate dispersion measure observed with pulsar and magnetar data at low frequencies, or with the fast radio bursts to lower the upper bound for photon mass using earth and space based telescopes. It might be noted here that presence of space based radio telescopes would be of particular importance in these studies since the ionosphere is opaque for frequencies less than its characteristic plasma frequency $\sim 1 - 10\,{\rm MHz}$. Gravitational lensing also is another method on constraining the photon mass because bending angle deviates depending on $m$ \cite{Okun:2006pn,Lowenthal:1973ka,Tu:2005ge,Spavieri:2011zz}.

Theoretical studies on implications of massive photons are also conducted to shed light on observational studies. For instance, it is found in \cite{Heeck:2013cfa} that even if the largest allowed value for the photon mass from other experiments is considered, the lower limit on the photon rest-frame lifetime is about 3 yr, which corresponds to a lifetime much larger than the age of the universe $10^{10}$ yr, around $10^{18}$ yr, for the photons in the visible spectrum. In the context of Standard Model Extensions, the Carroll-Field-Jackiw model \cite{Carroll:1989vb} is supersymmetrised, and four general classes of Super Symmetry (SuSy) and Lorentz Symmetry (LoSy) breaking were analysed \cite{Bonetti:2016vrq}. It is claimed in \cite{Bonetti:2016vrq} that the presence of photon mass may lead to observable imprints on dispersion measure at our energy scales. Meanwhile spin measurements for the largest known supermassive black holes could give an upper bound to $m \leq 10^{-22}$ eV \cite{Pani:2012vp}. Perhaps these new frontiers in astrophysics can be utilised in understanding of the microscopic universe. Indeed even a huge gap between the energy scales, quantum and cosmological scales intertwined to each other, e.g., particle creation that occur in the vicinity of Black Holes leads to Hawking radiation. The higher the scale the tighter the bounds shows us that even an extremely small value of the photon mass can have a considerable effect on the evolution of the universe.

\section{Scalar and vector field theories in cosmology}
\label{scalarvectorcosmo}
Scalar fields coupled either to gravity or to matter source are ubiquitous in particle physics inspired by unification theories like string theories. In cosmology, inflation models \cite{Starobinsky:1980te,Guth:1980zm,Albrecht:1982wi,Linde:1981mu} suggesting that the universe had gone through an accelerating expansion period at energy scales $\sim 10^{16}$ GeV just after the Big Bang are mostly based on the scalar field (inflaton) dynamics. Inflation mechanism is now part of the standard cosmology, since it does not only elegantly solve the problems of the standard big bang cosmology such as horizon and flatness problems but also generates the density fluctuations in the CMB (for a review see \cite{Linde:2007fr,Linde:2014nna}). A list of inflatons that have been studied can be found in \cite{Martin:2013tda}. The problem here is that there is no concrete realization of the inflatons from a fundamental theory such as string theory \cite{Quevedo:2002xw} and generally scalar fields used for describing inflaton are added to a model in an ad hoc way. 

There is no doubt today that the expansion of the universe started to accelerate once again approximately $6$ Gyr ago as confirmed with independent studies \cite{Riess:1998cb,Percival:2009xn,Bennett:2012zja,Ade:2013zuv}. This current acceleration that happens at energy scales $\sim 10^{-4}\,{\rm eV}$, which we are supposed to know the physics of this energy scale very well, is still a mystery. The simplest but yet very successful cosmological model accommodating this fact, we know so far, is the $\rm \Lambda$CDM model based on GR. It is the cosmological constant $\rm \Lambda$, which is mathematically equivalent to the conventional vacuum energy, what is responsible for the accelerating expansion in this model. However it has been well known that the $\rm \Lambda$ of this model suffers from two theoretical problems, so called fine tuning and coincidence problems, that are interpreted as a motivation for searching more general energy sources, collected under the name dark energy, for driving the current acceleration of the universe \cite{Zel'dovich:1968zz,Weinberg:1988cp,Sahni:1999gb,Peebles:2002gy,Copeland:2006wr,Bamba:2012cp}. Besides these theoretical problems, although there are observational reasons for questioning the $\rm \Lambda$ assumption. For instance, it is found in the Planck experiment that the data alone is compatible with $\rm \Lambda$ assumption, but when the astrophysical data is also taken into account \cite{Ade:2013zuv}, the dark energy yielding a time varying equation of state (EoS) parameter, which is typical for scalar fields, is favored. As for another recent example, the SDSS DR11 measurement of $H(z)=222\pm 7$ km/sec/Mpc at $z=2.34$ implies considerable tension with the $\rm \Lambda$CDM model \cite{Sahni:2014ooa,Aubourg:2014yra}: the overall flat $\rm \Lambda$CDM model fits the data well and its $\chi^2$ is tolerable but the Ly-$\alpha$ forest (LyaF) BAO measurements are in $(2-2.5 \sigma)$ tension with flat $\rm \Lambda$CDM predictions. Dynamical dark energy models are proposed in \cite{Sahni:2014ooa,Aubourg:2014yra} with a hope to decrease this inconsistency due the $\rm \Lambda$ assumption \cite{Delubac:2014aqe,Aubourg:2014yra}. It is common in cosmology to describe a dynamical dark energy by using scalar fields. However, the scalar field models of dark energy are also mostly ad hoc as in the inflation models or are motivated phenomenologically rather being derived from a fundamental theory (see \cite{Copeland:2006wr,Bamba:2012cp}) for comprehensive reviews on dark energy). 

Although the usage of scalar fields in cosmology are very common, the only known fundamental scalar field was the Higgs boson. Therefore, there have been attempts to relate inflation to the Higgs boson even before its discovery \cite{Bezrukov:2008ej}. However, Higgs boson was discovered \cite{Aad:2012tfa,Chatrchyan:2012xdj} with a mass $125\;{\rm GeV}$, which is not consistent with the typical energy scales of inflation. But yet, this discovery increased the interest in the the possible existence of scalar fields other than Higgs field with a mass consistent with the required energy scales for inflaton and dark energy fields. Rather than considering the presence of inflaton or dark energy sources by taking GR as the true of theory of gravity, the another way of obtaining accelerating expansion is to modify GR \cite{Nojiri:2010wj,Capozziello:2011et,Clifton:2011jh}. Scalar-tensor theories are the most established and well studied modified theories, and appear at low-energy limits of string theories. The prototype of such theories is the Brans-Dicke theory of gravity \cite{Brans:1961sx}, which involves an extra field mediating the gravitational interaction, namely, a scalar field that gives rise to a dynamical effective gravitational coupling by coupling directly to the scalar curvature. In GR on the other hand a scalar field is introduced as an external energy source (as an energy-momentum tensor) since GR is a pure tensor theory of gravity. Stueckelberg mechanism involving scalar field here becomes appealing since it gives mass to abelian vector fields and the Higgs field is still not considered as the inflaton candidate (though, we should note that recent works appear about the possible relation between the Higgs particle and inflation \cite{Hamada:2014iga,Bezrukov:2014bra}). The only remained mass mechanism, the Stueckelberg trick, remains to be used in cosmology. But we need an abelian vector field to piece the story together. 

Rather than gravity, the long range interaction which can be relevant on cosmological scales is the electromagnetic field, which is a vector field. The electromagnetic field is the radiation, whose energy density is inversely proportional to the fourth power of the scale factor, is typically not able to drive accelerated expansion in GR. In fact, scalar fields also are not able to give rise to accelerated expansion unless they are coupled to gravity non-minimally or have a potential. The possibility of inflationary models in which inflation is driven by a vector field rather than a scalar field is discussed by Ford \cite{Ford:1989me} at the end of 80's but started to receive keen attention only a decade ago. In recent years, on the other hand, vector fields have been discussed and considered with an increasing interest not only as an alternative to the scalar field models of inflaton but also that of dark energy \cite{Koivisto:2005mm,Dimopoulos:2007ns,Bamba:2008ja,Jimenez:2008au,Golovnev:2008cf,Koivisto:2008xf,Kanno:2008gn,Jimenez:2008nm,Watanabe:2009ct,Jimenez:2009sv,Kanno:2010nr,Golovnev:2009rm,Thorsrud:2012mu,Bartolo:2013msa}. Some of the anomalies found in the large-scale CMB temperature in the WMAP data \cite{WMAP7an} were the main reason behind this increase in interest to the vector fields and these anomalies have also been confirmed by the recent high precision Planck data \cite{Ade:2013zuv,Ade:2013nlj,Ade:2013vbw}. There is a large literature that argues that CMB multipole alignments, QSO polarisation alignment and large scale bulk flows all prefer approximately the same direction in the sky. Preference of a similar direction has recently been shown in the CMB maximum temperature asymmetry axis \cite{Mariano:2012ia} and in the direction dependence of the acceleration of the universe \cite{Antoniou:2010gw,Cai:2011xs,Zhao:2013yaa}. The Planck experiment also concludes that the most significant large-scale anomalies in the statistical isotropy of the CMB temperature such as the quadrupole-octopole alignment, hemi-spherical asymmetry represent real features of the CMB sky \cite{Ade:2013nlj}. These observed anomalies brought the isotropy assumption of the standard cosmology into question and increased the interest in vector fields that can yield anisotropic stress. For instance, it was showed that when the large-scale spatial geometry of the universe is allowed to be ellipsoidal with eccentricity at decoupling of order $10^{-2}$ that could be generated by a uniform cosmic magnetic field, the quadrupole problem can be solved \cite{Campanelli:2007qn}.
It is well-known that the generic inflation mechanism is based on scalar fields and predict an almost completely isotropic universe \cite{Wald:1983ky,Moss:1986ud,Kitada:1991ih}. Hence, if it is true that the space is actually anisotropic, then one should either introduce an inflationary scenario in which a small anisotropy could be maintained at the end of the inflation or introduce a mechanism that can anisotropize the universe slightly after the inflation took place. Considering anisotropic sources is the most obvious way of altering the isotropization dynamics. Analyses of cosmological evolution with known matter sources that possess small anisotropic pressures; electric/magnetic fields, simple topological defects, spatial curvature anisotropies and etc. are given in \cite{Barrow97}. However, because such sources should have been dominated by dust and then dark energy (DE) since the decoupling, in regard of the possibly slightly anisotropic geometry of the universe, the possibility of anisotropic models of inflaton and/or DE sources comes into question. Alternatively, generic scalar field inflation may be kept as it is and some anisotropy can be induced in relatively recent times relying on a DE source that yields an anisotropic EoS, e.g. a vector field, and hence accelerates the universe anisotropically \cite{Koivisto:2005mm,Koivisto:2008xf,Koivisto:2007bp,Rodrigues:2007ny,Battye:2009ze,Akarsu:2008fj,Akarsu:2010zm,Campanelli:2011uc,Thorsrud:2012mu,Akarsu:2013dva}. The possibility of anisotropic DE are subjects of current observational studies as well \cite{Appleby:2009za,Campanelli:2011uc,Appleby:2012as}. By default vector field may seem that directly brings direction dependency, however vector fields can be arranged to preserve the isotropy such as a triplet of orthogonal vector fields, known as the cosmic triads. On the other hand, for $N$ randomly oriented vector fields, the substantial anisotropy of the expansion of order $1/\sqrt{N}$ lives until the end of inflation \cite{Golovnev:2008cf}. A triplet of mutually orthogonal vector fields \cite{Bento:1992wy,ArmendarizPicon:2004pm} (see also \cite{Hosotani:1984wj,Galtsov:1991un} for exact isotropic solutions of the Einstein-Yang-Mills system based on the same idea) or a large number of randomly oriented fields can be used. Another possibility is to consider purely time-like vector fields \cite{Jimenez:2008au,Koivisto:2008xf,Kiselev:2004py,Carroll:2004ai,Boehmer:2007qa}. It is discussed in Ref. \cite{Jimenez:2008au} that such vector theories of dark energy can additionally offer a relief from fine tuning problems, for which not only the cosmological constant but also dynamical dark energy models that are usually based on scalar fields or modified theories of gravity said to be suffering from.

However, vector field models that give an accelerated expansion usually suffer from ghost instabilities \cite{Himmetoglu:2008zp,Himmetoglu:2009qi,EspositoFarese:2009aj}, namely, such models require vector fields with imaginary (tachyonic) mass to be able to give rise to accelerating expansion. Moreover vector inflationary models require huge mass for the vector field, and hence a huge amount of tachyonic mass which makes the issue even worse. This shows us the mass of the vector field is problematic and it needs a deeper thinking about the mass notion of the vector field. May be the appropriate way to give the mass term is not the Proca way and different mechanisms should be searched so this is another motivation is to apply Stueckelberg theory to cosmology. There has recently been great interest in inflationary cosmologies in which a canonical scalar field is non-minimally coupled to the vector kinetic term \cite{Watanabe:2009ct,Kanno:2010nr} as
 \be
 \frac{1}{2}f(\phi)^2F_{\mu\nu}F^{\mu\nu}
 \ee
that often appears in supergravity \cite{Martin:2007ue} and the instability problems do not arise \cite{Watanabe:2009ct,Kanno:2010nr,Hervik:2011xm,Fleury:2014qfa}. In such models, the vector field however appears as the spectator field, namely, it is the scalar field that drives the accelerated expansion, while the vector field, which becomes persistent due to its non-minimal coupling, leads to a non-vanishing expansion anisotropy. On the other hand, in the absence of sources like cosmological constant, scalar field etc., it is possible to find viable accelerating cosmologies in the framework of gravity theories constructed by non-minimal coupling of the vector field to gravity. For instance, in a recent series of papers, the generalisation of the Proca action for a massive vector field with derivative self-interactions has been constructed \cite{Heisenberg:2014rta,Jimenez2016isa} systematically so as not to allow any instabilities, which requires non-minimal gravity coupling of vector field when it is extended to a curved spacetime, and its cosmological significance, particularly in the context of dark energy, has been studied \cite{DeFelice:2016uil,DeFelice:2016yws,Heisenberg:2016wtr}. Scalar and vector fields, minimally or non-minimally coupled either to gravity or each other, have found an extensive usage in cosmology to tackle with some important issues. Leading to the presence of scalar field in a natural way to give a mass to a vector field by preserving gauge invariance makes Stueckelberg mechanism appealing for considering it in the context of cosmology. Because, if photons in the universe are indeed massive and their mass is generated via the Stueckelberg mechanism, then the presence of photons implies the presence of a scalar field other than Higgs boson in the universe. It is known that neither a vector field nor a scalar field without a suitable potential are able to give rise to accelerated expansion of the universe, hence it would be wise to couple Stueckelberg fields to gravity non-minimally, like in the Brans-Dicke theory of gravity, which is the prototype of scalar-tensor gravity theories, and preserve the gauge-invariant feature of the Stueckelberg mechanism. In what follows we will give a cosmological application of the Stueckelberg mechanism in accordance with all these points we discussed.

\section{A cosmological model using Stueckelberg mechanism}
\label{stucurved}
In a recent work \cite{Akarsu:2014eaa}, we coupled the gauge fixing term of the Stueckelberg Lagrangian density \eqref{procastu} to the gravity, namely, to the scalar curvature, and obtained an interesting solution for which the universe exhibits a decelerating expansion phase sandwiched by two different accelerated expansion phases.

 The Lagrangian density we considered in the original model \cite{Akarsu:2014eaa} is in the following form:
\bea
 {\cal L}=-\frac{1}{8\omega m^2}\left(\nabla_{\mu} A^{\mu}+m B\right)^2 R+ {\cal L}_{\rm Stu}+ {\cal L}_{\rm m},
\label{actions}
\eea
where $\omega$ is a dimensionless coupling constant, $R$ is the scalar curvature of the spacetime metric $g$, respectively. Here the mass term $m$ is defined as a real valued positive number, so that we do not allow imaginary (tachyonic) mass for the vector field that leads to ghost instability \cite{Himmetoglu:2008zp,Himmetoglu:2009qi,EspositoFarese:2009aj}. The ${\cal L}_{\rm m}$ stands for the Lagrangian density of the matter source. We note that the reduced Planck mass is given by $M_{\rm pl}=1/\sqrt{8\pi G}$, where $G$ is the gravitational coupling, in line with our natural units with $\hbar=c=1$ we mentioned above. Hence, defining
\be
\label{eqn:ff}
f=\nabla_{\mu} A^{\mu}+mB,
\ee
for convenience, we see that $f$ is related to the gravitational coupling parameter as
\bea
\frac{f^2}{8\omega m^2} = 16\pi G
\label{fG}
\eea
when compared to the Einstein-Hilbert action of the standard GR. Hence, the dynamics of the Stueckelberg fields (coupled vector and scalar fields) will determine the function $f$ and hence can lead a dynamical effective gravitational coupling. The other side of the coin is that one can consider to interpret this model, the Lagrangian density we introduced, as a modified Stueckelberg Lagrangian density such that the gauge fixing term of the conventional Stueckelberg is rescaled with a scalar curvature $R$ dependent function which may also be time dependent.

It might be useful at this point to comment on whether the model by construction leads to instabilities or not. To do so, we first overview Hamiltonian of some well known systems that are relevant to the model under discussion here. In Maxwell theory, Hamiltonian is unbounded below, but a perturbative analysis does not reveal any instabilities. The reason behind this is that there are fewer propagating degrees of freedom at the linear level (a spin-1 mode propagates, but not a spin-0 mode). In Proca theory the Hamiltonian is already positive definite owing to Lorentz subsidiary condition as mentioned in section \ref{intro} and therefore there is no instability. In the standard Stueckelberg theory, on the other hand, like in QED, the Lorentz subsidiary condition is not automatically derived. However, Stueckelberg imposes the physical matrix elements of the gauge fixing term $f$ to be zero so that he could remove the unphysical states, i.e., the spin-0 mode does not propagate and Hamiltonian becomes bounded below, relying on the presence of scalar field in $f$. Let us now direct our attention to the gravitational sector. In the standard GR, the Einstein-Hilbert action includes second derivatives of the metric and its Hamiltonian is linear in at least one of its canonical momenta that signal Ostrogradski's instability \cite{ostro1850}. However it avoids this instability since it does not satisfy the non-degeneracy assumption of Ostrogradski's instability. In modified theories of gravity on the other hand the presence of higher-order derivatives or giving up diffeomorphism invariance may lead to more degrees of freedom and these models may suffer from Ostrogradski's instability. However, it is still possible to save such theories, e.g., $f(R)$ theories, from Ostrogradski's instability using some methods (partial integration, gauge invariance, etc.). The presence of such instabilities is a complication that may arise in modified theories of gravity involving an extra degree(s) of freedom non-minimally coupled to gravity and hence should be investigated carefully \cite{modifiedgr}. In the model under consideration here \eqref{actions}, on the other hand, the Stueckelberg and the Einstein-Hilbert Lagrangians, which are known to be stable as aforementioned above, are non-minimally coupled without introducing any new degrees of freedom, such that we couple $f$ to $R/8\omega m^2$ as $fR$ in curved spacetime so that the gauge fixing term is promoted to a dynamical term but yet is still the gauge fixing term that does not affect the dynamical degrees of freedom, and therefore we do not expect Ostrogradski's instability. However, thorough analysis of the stability of this model or its proof are out of scope of this mini review.

Varying the action \eqref{actions} with respect to the inverse metric we obtain the following Einstein field equations
\begin{eqnarray}
\label{einsteintensor}
&& \frac{f^2}{4\omega m^2}G_{\mu\nu}-\frac{1}{2}g_{\mu\nu}f^2+\frac{1}{4\omega m^2}(g_{\mu\nu}\Box-\nabla_{\mu}
\nabla_{\nu})f^2+\bigg(\frac{R}{4\omega m^2}+1\bigg)(-2\nabla_{(\mu}fA_{\nu)}\nonumber \\
&&+\nabla^{\alpha}A_{\alpha}g_{\mu\nu}f+g_{\mu\nu}\nabla^{\alpha}fA_{\alpha})
-\frac{f}{4\omega m^2}\left(2A_{(\nu}\nabla_{\mu)}R+g_{\mu\nu}A_{\alpha}\nabla^{\alpha}R\right)+ F_\mu ^a F_{\nu a} \nonumber \\
&&-\frac{1}{4}g_{\mu\nu}F^{\alpha\beta}F_{\alpha\beta}-(\partial_{\mu}B-mA_{\mu})(\partial_{\nu}B-mA_{\nu}) +\frac{1}{2}g_{\mu\nu}(\partial_{\alpha}B-mA_{\alpha})^2=T_{\mu\nu}^{({\rm M})}, \nonumber \\
\end{eqnarray}
where $T_{\mu\nu}^{({\rm M})}$ is the energy-momentum tensor of the matter source. The variations of \eqref{actions} with respect to the vector field $A_{\mu}$ yield the vector field equation
\begin{eqnarray}
\label{vectoreq}
\nabla_{\alpha}F^{\alpha\mu}+\nabla^{\mu}(\nabla_{\alpha}A^{\alpha})+m^2A^{\mu}+\frac{1}{4\omega m^2}\left(f\nabla^{\mu}R+R\nabla^{\mu}f\right)=0,
\end{eqnarray}
and with respect to $B$ give
\begin{equation}
\label{scalareqq}
\left(\Box+m^2\right)B+\frac{R f}{4\omega m}=0.
\end{equation}
In what follows, we will use the following notation
\be
A_{\mu}=(A_0,A_1,A_2,A_3)=(A,A_x,A_y, A_z).
\ee
 
\subsection{A solution in spatially flat Robertson-Walker background}
\label{isotropic}
We consider, for convenience, the spatially maximally symmetric spacetime metric, i.e., Robertson-Walker metric, with flat space-like sections 
\begin{eqnarray}
\label{eqn:rw}
{\rm d}s^2={\rm d}t^2-a^2[{\rm d}x^2+{\rm d}y^2+{\rm d}z^2],
\end{eqnarray}
where $a=a(t)$ is function of cosmic time only. We assume the space-like components of the vector field are zero in line with the isotropy property of the metric \eqref{eqn:rw},
\be
A_{\mu}=(A,0,0,0),
\ee 
where $A=A(t)$ is the function of cosmic time only. Considering this metric, the field equations read
\begin{eqnarray}
 && \frac{3f^2}{4\omega m^2}\left(\frac{\dot{a}^{2}}{a^{2}}\right)-\frac{f^2}{2}+\left(\frac{R}{4\omega m^2}+1\right)\left(-\dot fA+f\dot A+3fA \frac{\dot{a}}{a}\right)
 +\frac{3f \dot f}{2\omega m^2}\frac{\dot{a}}{a} \nonumber \\
 &&-\frac{fA\dot{R}}{4\omega m^2}-\frac{1}{2}(\dot B-mA)^2=\rho^{({\rm M})},
\label{rho}
\end{eqnarray}
\begin{eqnarray}
&&-\frac{f^2}{4\omega m^2}\left(2\frac{\ddot{
a}}{a}+\frac{\dot{a}^{2}}{a^{2}}\right)+\frac{f^2}{2}-\frac{1}{4\omega m^2}\left(2\dot f^2+2f\ddot f+4f\dot{f}\frac{\dot{a}}{a}\right) -\frac{fA\dot{R}}{4\omega m^2} \nonumber \\
&&-\left(\frac{R}{4\omega m^2}+1\right)\left(\dot fA+f\dot A+3fA\frac{\dot{a}}{a}\right)-\frac{1}{2}(\dot B-mA)^2=p^{({\rm M})},
\label{pressure}
\end{eqnarray}
\begin{eqnarray}
\ddot A+3\dot A\frac{\dot{a}}{a} +3A\left(\frac{\ddot{
a}}{a}-\frac{\dot{a}^{2}}{a^{2}}\right)+m^2 A+\frac{\dot R f+\dot{f}R}{4\omega m^2}=0,
\label{vector}
\end{eqnarray}
\begin{eqnarray}
\ddot B+3 \dot B \frac{\dot{a}}{a}+m^2 B+\frac{Rf}{4 \omega m}=0,
\label{scalar}
\end{eqnarray}
where
\begin{eqnarray}
\label{notaconstant}
f=mB+\dot{A}+3 A \frac{\dot{a}}{a},
\end{eqnarray}
and $\rho^{({\rm M})}$ and $p^{({\rm M})}$ are the energy density and the pressure of the matter source.

The system has four linearly independent ordinary differential equations \eqref{rho}-\eqref{scalar} and five unknown variables, $\rho^{\rm (M)}$, $p^{\rm (M)}$, $A$, $B$ and $a$, thus is not fully determined. It is common at this point to define a cosmologically relevant equation of state (EoS), $w=p/\rho$, for the matter source (radiation, dust, cosmological constant) to close the system. Because the system is far too complicated to solve analytically, in \cite{Akarsu:2014eaa}, not just for convenience but also for setting the effective gravitational coupling to a constant, solutions for this system have been looked for by assuming $f={\rm constant}$ utilising the relation between $f$ and the effective gravitational coupling $G$ given in \eqref{fG}. One may see the original paper \cite{Akarsu:2014eaa} for the details of the case $f=0$ and discussions about the cases $f={\rm constant}\neq0$. Here, in this mini review, we basically proceed with the case $f=0$, which gives rise to a universe that goes through a decelerating expansion phase sandwiched by two different accelerated expansion phases.

We content ourselves here with the particular case $f=0$ of this solution. Setting $f=0$ equations \eqref{rho}-\eqref{scalar} reduce to the following
\begin{eqnarray}
\label{rhof0}
-\frac{1}{2}(\dot B-m A)^2=\rho^{({\rm m})},\\
\label{p0}
-\frac{1}{2}(\dot B-m A)^2=p^{({\rm m})},\\
\label{vecf0}
\ddot A+3\dot{A} \frac{\dot{a}}{a} +3A \left(\frac{\ddot{a}}{a}-\frac{\dot{a}^2}{a^2}\right)+m^2 A=0,\\
\label{sclr0}
\ddot B+3\dot{B} \frac{\dot{a}}{a}+m^2 B=0,
\end{eqnarray}
and
\be
\label{notaconstant1}
mB+\dot{A}+3 A \frac{\dot{a}}{a}=0.
\ee
When the derivative of \eqref{notaconstant1} is used in \eqref{vecf0}, we obtain $A=\dot{B}/m$ and substituting this into \eqref{notaconstant1} we get \eqref{sclr0}, implying that it is linearly dependent to \eqref{notaconstant1} and \eqref{vecf0} and hence the solution of \eqref{notaconstant1} and \eqref{vecf0} together will always satisfy \eqref{sclr0}. We note in this particular case $f=0$ that $A=\dot{B}/m$ implies $\rho^{({\rm M})}=p^{({\rm M})}=0$, that is the system does not permit any matter source such as radiation, dust and hence only the Stueckelberg fields play role on the evolution of the universe. Eventually there left two linearly independent equations but three unknown functions $A$, $B$ and $a$ and hence we are free to introduce one more equation to close the system.

If we think of a play between the scalar and vector fields as the universe evolves, then it might be useful to define a ratio as $F = A/B$, which might provide us with an insight for choosing a useful and reasonable function for the additional constraint rather than an arbitrary function (see \cite{Akarsu:2014eaa} for the details): Equation \eqref{notaconstant1} gives the scale factor as 
\begin{equation}
\label{eqn:inf1}
a=a_1 e^{-\frac{1}{3}\int m\frac{B}{A}+\frac{\dot{A}}{A}\;{\rm d}t},
\end{equation}
where $a_1$ is an integration constant. Using $A=\dot{B}/m$ and $F = A/B$ together with \eqref{eqn:inf1}, we see that scale factor, and consequently Hubble parameter, includes three additive terms;
\begin{equation}
\label{eqn:f0gen}
a=a_1 e^{-\frac{1}{3} \int \frac{m}{F}+\frac{\dot{F}}{F}+mF\; {\rm d}t} \quad\textnormal{and}\quad H=-\frac{1}{3}\frac{m}{F}-\frac{\dot{F}}{F}-mF.
\end{equation}
Just looking at the mathematical form of the Hubble parameter we obtained here, it is easy to see that if $m\neq0$ then even a simple power-law assumption for $F$ can lead to an expansion history that involves characteristically different three epochs depending on the sign of $F$ as well as the sign and the value of the power, while such an assumption would lead nothing but a simple power-law expansion in case $m=0$. Accordingly we choose that $F$ obeys a power-law ansatz as 
\begin{equation}
\label{eqn:asmpf0}
F=\frac{\mathcal{A}_0}{\mathcal B_0} \left(\frac{t}{t_0}\right)^{-k},
\end{equation}
where $t_0$, $\mathcal{A}_0$ are $\mathcal{B}_0$ constants, and $k$ is another constant whose sign will also determine whether the vector field will be dominant over the scalar field at the earlier times or the later times. Solving the system using this constraint the scale factor reads
\begin{eqnarray}
\nonumber
 a=a_{0}t^{\frac{k}{3}}{\rm e}^{\frac{\mathcal{A}_{0}}{\mathcal B_{0}}\frac{m t_{0}}{3(k-1)} \left(\frac{t}{t_{0}}\right)^{-k+1}}{\rm e}^{-\frac{\mathcal B_{0}}{\mathcal A_{0}}\frac{m t_{0}}{3(k+1)}\left(\frac{t}{t_{0}}\right)^{k+1}} \quad &\textnormal{for}& \quad |k| \neq 1,
\\
\label{eqn:infsol}
a = a_0 t^{-\frac{\mathcal B_{0}}{\mathcal A_{0}}\frac{mt_{0}}{3}-\frac{1}{3}} {\rm e}^{-\frac{\mathcal A_{0}}{\mathcal B_{0}}\frac{mt_{0}}{6}\left(\frac{t}{t_{0}}\right)^2}\quad &\textnormal{for}& \quad k=-1,
\\
\nonumber
a = a_0 t^{-\frac{\mathcal A_{0}}{\mathcal B_{0}}\frac{mt_{0}}{3}+\frac{1}{3}} {\rm e}^{-\frac{\mathcal B_{0}}{\mathcal A_{0}}\frac{mt_{0}}{6}\left(\frac{t}{t_{0}}\right)^2}\quad &\textnormal{for}& \quad k=1,
\end{eqnarray}
where $a_0$ is an integration constant.

The corresponding Hubble parameter $H=\frac{\dot{a}}{a}$ and deceleration parameter $q=\frac{{\rm d} }{{\rm d} t}\left(\frac{1}{H}\right)-1$ for this solution are as follows:
\begin{equation}
\label{eqn:hinf}
H=-\frac{\mathcal A_{0}}{\mathcal B_{0}} \frac{m}{3} \left(\frac{t}{t_{0}}\right)^{-k} +\frac{k}{3}t^{-1}  -\frac{\mathcal B_{0}}{\mathcal A_{0}}\frac{m}{3}  \left(\frac{t}{t_{0}}\right)^{k},
\end{equation}
and
\begin{equation}
\label{eqn:qinf}
q=3k{t_{0}}^{k}t^{k-1}\frac{\mathcal A_{0}}{\mathcal B_{0}} \frac{\frac{\mathcal A_{0}}{\mathcal B_{0}} {t_{0}}^{k}  t^{k-1} + m(t^{2k}-\frac{{\mathcal A_{0}}^{2}}{{\mathcal B_{0}}^{2}} {t_{0}}^{2k})}{\left[  k \frac{\mathcal A_{0}}{\mathcal B_{0}} {t_{0}}^{k}  t^{k-1} - m(t^{2k}+\frac{{\mathcal A_{0}}^{2}}{{\mathcal B_{0}}^{2}} {t_{0}}^{2k}) \right]^2}-1.
\end{equation}
The universe exhibits accelerated expansion if $H>0$ and $q<0$. One may check that the model gives rise to various behaviors depending on the choice of the parameters. Under the assumption $\mathcal A_{0}/\mathcal B_{0}<0$ and $k>1$ the universe starts expanding at $t=0$ and will always expand passing through three different stages respectively:
\paragraph{Early accelerated phase with a large expansion rate (Inflationary phase):} We note that \eqref{eqn:infsol}-\eqref{eqn:qinf} approximates as follows
\begin{eqnarray}
\label{phaseI}
 a&\sim &\exp\left[\frac{\mathcal A_{0}}{\mathcal B_{0}}\frac{m t_{0}}{3(k-1)} \left(\frac{t}{t_{0}}\right)^{-k+1}\right], \\
 \nonumber
 H&\sim &-\frac{\mathcal A_{0}}{\mathcal B_{0}} \frac{m}{3} \left(\frac{t}{t_{0}}\right)^{-k}
 \quad
 q\sim -\frac{3k}{mt_{0}}\frac{\mathcal B_{0}}{\mathcal A_{0}}\left(\frac{t}{t_{0}}\right)^{(k-1)}-1\quad \textnormal{at} \quad t\simeq 0.
\end{eqnarray}
 According to this, the universe begins with an accelerating expansion rate, such that $a\rightarrow 0$, $H\rightarrow\infty$ and $q\rightarrow -1$ as $t\rightarrow 0$. The $H$ can be set so as to get sufficiently large values for the time when inflation took place by choosing properly large values for the ratio $|\mathcal A_0/\mathcal B_0|$ and holding the mass term $m$ at low values consistent with the observational constraints on it.

\paragraph{Decelerated expansion phase:} Provided that the parameters are chosen properly, the power term will be dominant over the two exponential terms in \eqref{eqn:infsol} after a while and the second phase in which the evolution of the universe can be approximately described by \eqref{phaseII} will start:
\begin{equation}
\label{phaseII}
 a\sim t^{\frac{k}{3}},
 \quad
 H\sim \frac{k}{3}t^{-1}
 \quad\textnormal{and}\quad
 q\sim \frac{3}{k}-1\quad \textnormal{at}\quad t \gtrsim 0.
\end{equation}
In this phase \eqref{phaseII}, provided that $0<k<3$, the universe exhibits decelerating expansion, i.e., $H>0$ and $q>0$. Thus accelerated expansion in the previous phase (in case $k>1$) can end, namely, the inflation can be switched off, and the universe can enter into the decelerated expansion phase.

\begin{figure}[t]\centering
\includegraphics[width=0.7\textwidth]{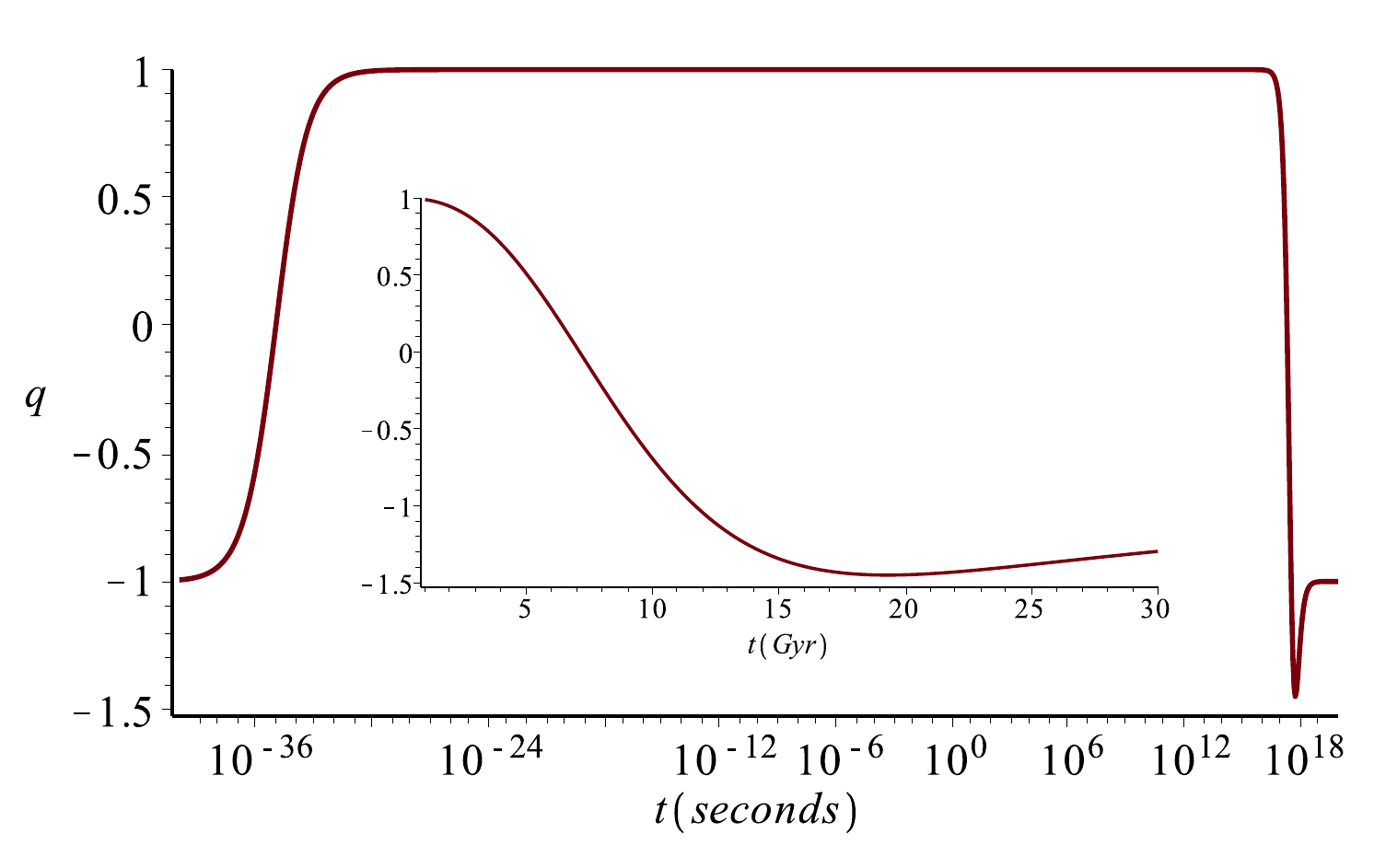}
\caption{Deceleration parameter $q$ versus cosmic time $t$.}
\label{fig:qvst}
\end{figure}

\paragraph{Late-time acceleration and a low expansion rate:} Eventually, the last exponential term will be dominant over the exponential term at the middle and the power term in \eqref{eqn:infsol} and the third phase, in which the universe will be described by \eqref{phaseIII}, will start,
\begin{eqnarray}
\label{phaseIII}
a&\sim &\exp\left[-\frac{\mathcal B_{0}}{\mathcal A_{0}}\frac{m t_{0}}{3(k+1)}\left(\frac{t}{t_{0}}\right)^{k+1}\right], \\
\nonumber
 H&\sim& -\frac{\mathcal B_{0}}{\mathcal A_{0}}\frac{m}{3}  \left(\frac{t}{t_{0}}\right)^{k} \quad
 q\sim \frac{3k}{mt_{0}}\frac{\mathcal A_{0}}{\mathcal B_{0}}   \left( \frac{t}{t_{0}}\right)^{-k-1}-1\quad \textnormal{at}\quad t\gg0.
\end{eqnarray}
We note that the conditions $\mathcal A_{0}/\mathcal B_{0}<0$ and $3>k>1$ we considered lead the deceleration parameter given in \eqref{phaseIII} to be less than $-1$ and $q\rightarrow-1$ as 
$t\rightarrow\infty$. Hence as the contribution of the power term given \eqref{phaseII} weakens with respect to exponential term given in \eqref{phaseIII} as the universe expands and the deceleration parameter will eventually evolves into negative values and the universe will start to accelerate once again. But this time with an expansion rate much smaller than the expansion rate of the first accelerated expansion phase that took place just after the Big Bang: We note that we have the factor $|\mathcal A_0/\mathcal B_0|$ in the Hubble parameter in the inflation phase \eqref{phaseI}, while the factor is its inverse as $|\mathcal B_0/\mathcal A_0|$ in the late time acceleration phase \eqref{phaseIII}. According to this, properly chosen large values of $|\mathcal A_0/\mathcal B_0|$ for obtaining larger $H$ values in line with inflation scenario lead automatically to low values for $H$ for the late time acceleration of the universe. Hence the ratio between Hubble parameters of inflation phase to late time acceleration phase is the time evolution of the the Stueckelberg vector field to scalar field ratio, given as $H_{\rm inflation}/H_{\rm today}\sim F^2=(A/B)^2$. This ratio demonstrates the role of the combining Stueckelberg fields explicitly and, maybe, the reason of such a large difference in the expansion rates of the early and current universe. It is remarkable that this whole interesting scenario is a result of the interplay between the inseparable Stueckelberg fields. Whether this such an intriguing and natural scenario which describes the evolution of the universe in the correct order survive or not is a question for a deviation from $f=0$.
We demonstrate the behaviour of the model through the deceleration parameter $q$ in Figure \ref{fig:qvst} by giving some suitable values to the parameters. We first choose $k=\frac{3}{2}$, so that in the decelerated expansion phase the deceleration parameter becomes $q=1$, which is the value of the deceleration parameter of the radiation dominated universe, as well as at the time when Big Bang nucleosynthesis (BBN) took place ($\sim 10^2$ seconds after the Big Bang), in the standard cosmology based on GR. We choose $m=10^{-45}$ eV, which is a value almost 20 orders of magnitude less than the observational upper limits, and take the $H_{0}=10^{-32}$ eV and $\frac{A_{0}}{B_{0}}=-10^{-13}$ for today, when the age of the universe is $t_{0}=14\,{\rm Gyr}\simeq4.4\times10^{17}\,{\rm s}$.

Using these values we find that there is inflation in the very early universe, namely, $q\simeq -1$, $H\gtrsim 10^{39}\;{\rm s}^{-1}$ and $A/B\lesssim-10^{70}$ when $t\lesssim10^{-38}\;{\rm s}$, and it ends at about $t=10^{-35}\;{\rm s}$ such that $q\simeq0$, $H\simeq 10^{35}\;{\rm s}^{-1}$ and $A/B=-10^{66}$ at that time. In a short while following the end of the inflationary phase, the universe achieves an expansion rate with a deceleration parameter equal to unity and preserves this value for a long time, namely, till the age of the universe reaches $t\simeq 10^{17}\;{\rm s}$. Afterwards, the deceleration parameter starts to decrease, passes through $q=1/2$, which is the value of deceleration parameter of the matter dominated universe in GR, and becomes zero $q=0$ at $t\simeq 7\,{\rm Gyr}\sim2\times10^{17}\,{\rm s}$ and the universe starts to accelerate for the second time. In this late time acceleration period, the universe reaches even to a super-exponential expansion rate $q<-1$ and then, for the $t\gtrsim 10^{18}\,{\rm s}$, the deceleration parameter approaches monotonically to $q=-1$ as $t$ increases.

Let us now comment on several concerns that should be dealt to be able to upgrade our model to a more complete and realistic cosmological model. It is quite interesting that the pattern of expansion history (see Figure \ref{fig:qvst}) we obtained from our simple assumption \eqref{eqn:asmpf0} is kinematically in line with the one in our current paradigm of cosmology based on the inflation+$\rm \Lambda$CDM models. On the other hand, considering the latest precision data from observations, we know that in a successful cosmological model the expansion history of the universe should not deviate a lot from the one described in the current paradigm of cosmology. Hence, keeping this in mind, the expansion history of the universe in our model can be fine tuned by modifying or changing our power-law assumption on $F$ given in \eqref{eqn:asmpf0}. So that, one can, in principle, preserve the main pattern we give in Figure \ref{fig:qvst} for $q$ but fine tune it to be more realistic, namely, $q$ would take precisely correct values at the relevant time scales. For instance, it is expected in the standard cosmology that the age of the universe is $\sim10^5\,{\rm yr}$ and the deceleration parameter is equal to $\frac{1}{2}$ when the photon decoupling took place (CMB released), while our model predicts quite different value $q=1$ for that time. However, avoiding this mismatch, in principle, is just a matter of replacing our power-law assumption on $F$ with a suitable function. On the other hand, even if we could construct the evolution of $q$ in $t$ with a great success by a suitable assumption on $F$, one should recall that the choice $f=0$ in this solution implies infinitely large effective gravitational coupling and hence extending the model for large $t$ values may not be reliable. However, because there is no matter source ($\rho^{(\rm m)}=0$) in this solution, extending the model to large $t$ values will still be consistent within the model itself. On the other hand, once we introduce matter source $f$ shouldn't be null anymore, but either a non-zero constant or a dynamical function that varies sufficiently slowly, not to spoil the physical processes such as BBN, after a certain age of the universe. This point of the model deserves further discussion. Recall that the $f$ is not a true constant of our model but a dynamical parameter, given in \eqref{notaconstant},
consisting of three additive terms that are dynamical too. Hence, the investigation of a solution under the assumption $f=0$ should be understood as the investigation of the behaviour of our model in the period of time when the constituents of $f$ possibly evolve such that $f$ vanishes. The energy scale (so the time scales between the Planck time scale $10^{-43}$ s and SUSY breaking time scale $<10^{-10}$ s) is extremely high and beyond our technology that we have reached, the energy scale corresponding to up to $10^{?10}$ s, at the Large Hadron Collider (LHC). Therefore the gravitational coupling at time scales close to the Planck time scales may be large and this extreme case solution may be interpreted as the dynamics of the early universe, in the context of inflation believed to took place at time scales $\sim 10^{-35}$ s with the corresponding energy scales $\sim10^{15}$ GeV. Solutions giving rise to such an interesting behavior of the universe but not suffering from this issue may be obtained by allowing the effective gravitational coupling $G=\frac{\omega m^2}{2\pi f^2}$ to be a particular function of time such that it will start with infinitely large values but will then approach to a non-zero value by changing slowly enough after the end of inflation compatible with the observational constraints. For instance, it might be interesting to study $f^2\propto\frac{1}{G}\propto \tanh^{r}(t/t_0)$, where $r>0$. In such models the predictions on time variation of the fundamental constants should not be in conflict with the observational upper bounds and an observational constraint on the rate of change of the gravitational coupling today is $\mid\frac{\dot{G}}{G}\mid \leq 10^{-10}-10^{-12}$ 1/yr (see \cite{Uzan:2010pm} for a recent review). We may alternatively think of $f$ is being zero through the inflationary period but then becomes a non-zero constant spontaneously at some point just after inflation. Such a scenario can also be promising, because, although we didn't discuss it here in this mini review, one may see in \cite{Akarsu:2014eaa} that in case $f={\rm constant}\neq0$ the model under consideration not only already predicts the presence of radiation source in the universe but also a scale factor as $a\propto\sinh^{1/2}{\left(2\sqrt{\frac{w}{3}}mt\right)}$, which gives $a\propto t^{\frac{1}{2}}$ yielding $q=1$, as in case the radiation dominated universe in GR, for small $t$ values and accelerating expansion as $a\sim e^{\sqrt{\frac{\omega}{3}}mt}$ yielding $q\sim-1$ for large $t$ values. However, in this scenario, where $f$ becomes a non-zero constant spontaneously, pressure-less matter source remains as the missing part for a complete and realistic cosmology.
 
\subsection{Extensions to anisotropic spacetimes}
\label{anisotropic}
Let us now proceed with a brief discussion of the model when the space-like component of the vector field is considered. To do so, we first consider an electromagnetic four-potential yielding a magnetic potential along the $z$-axis as
\be
\label{eqn:a000zN}
A_{\mu}=(0,0,0,A_z),
\ee
and, so as to accommodate the anisotropic stress that arises from this choice, the simplest anisotropic generalisation of the spatially flat RW spacetime, namely, the spatially flat and homogeneous but not necessarily isotropic locally rotationally symmetric (LRS) Bianchi type-I spacetime metric as
\begin{eqnarray}
\label{bti}
{\rm d}s^2={\rm d}t^2- a^2[{\rm d}x^2+{\rm d}y^2]-b^2{\rm d}z^2,
\end{eqnarray}
where $a=a(t)$ and $b=b(t)$ are the directional scale factors along the $x$, $y$-axes and $z$-axis, respectively. In this case, the field equations \eqref{einsteintensor}-\eqref{scalareqq} read:
\begin{eqnarray}
\frac{\dot{a}^{2}}{a^{2}}+2\frac{\dot{a}}{a}\frac{\dot{b}}{b}+\frac{2\dot{B}}{B}\left(2\frac{\dot{a}}{a}+\frac{\dot{b}}{b}\right)-2\omega\frac{\dot{B}^2}{B^2}=\frac{4\omega}{B^2}\bigg\{ \frac{1}{2}\frac{\dot{A_z}^2}{b^2}+\frac{1}{2}m^2\left(B^2+\frac{A_z^2}{b^2}\right) \bigg\}, \nonumber \\
\label{rhoa} 
\end{eqnarray}
\begin{eqnarray}
\frac{\ddot{a}}{a}+\frac{\ddot{b}}{b}+\frac{\dot{a}}{a}\frac{\dot{b}}{b}+2\frac{\dot{B}}{B}\left(\frac{\dot{a}}{a}+\frac{\dot{b}}{b}\right)+2\frac{\ddot{B}}{B}+(2\omega+2)\frac{\dot{B}^2}{B^2}=
\frac{4\omega}{B^2}\bigg\{-\frac{1}{2}\frac{\dot{A_z}^2}{b^2} \nonumber \\
+\frac{1}{2}m^2\left(B^2+\frac{A_z^2}{b^2}\right)\bigg\},
\label{pressurex}
\end{eqnarray}
\begin{eqnarray}
2\frac{\ddot{a}}{a}+\frac{\dot{a}^2}{a^2}+4\frac{\dot{a}}{a}\frac{\dot{B}}{B}+2\frac{\ddot{B}}{B}+(2\omega+2) \frac{\dot{B}^2}{B^2}=
\frac{4\omega}{B^2}\bigg\{\frac{1}{2}\frac{\dot{A_z}^2}{b^2}+\frac{1}{2}m^2\left(B^2-\frac{A_z^2}{b^2}\right)\bigg\}, \nonumber \\
\label{pressurez}
\end{eqnarray}
\begin{eqnarray}
\frac{\ddot A_z}{A_z}+\left(2\frac{\dot{a}}{a}-\frac{\dot{b}}{b}\right)\frac{\dot A_z}{A_z}+m^2=0,
\label{vector}
\end{eqnarray}
\begin{eqnarray}
\ddot B+\left(2\frac{\dot{a}}{a}+\frac{\dot{b}}{b}\right)\dot B+m^2B+\frac{RB}{4\omega}=0,
\label{scalardif}
\end{eqnarray}
\begin{eqnarray}
\label{RR}
\frac{\dot{R}}{R}+\frac{\dot{B}}{B}=0,
\end{eqnarray}
 \begin{eqnarray}
 \label{toget}
T_{0i} =-m\dot B A_z=0,
 \end{eqnarray} 
 where $R=-4\frac{\ddot{a}}{a}-2\frac{\ddot{b}}{b}-4\frac{\dot{a}\dot{b}}{ab}-2\frac{\dot{a}^2}{a^2}$. From \eqref{toget} we see that the scalar field should be constant, $B={\rm const.}$, for non-zero the electric field $A_z\neq0$ and photon mass $m\neq0$. Substituting constant scalar field, $\dot B=0$, in \eqref{RR} we find 
\begin{eqnarray}
 B=-\frac{\kappa}{R},
 \end{eqnarray}
 where $\kappa$ is an integration constant, and in \eqref{scalardif} we get 
\begin{eqnarray}
R=-4\omega m^2.
\label{Rdef}
\end{eqnarray}
The model now can be described by the following set of three equations:
 \begin{eqnarray}
\frac{\dot{a}^{2}}{a^{2}}+2\frac{\dot{a}}{a}\frac{\dot{b}}{b}=8\pi G \left(\frac{1}{2}\frac{\dot{A_z}^2}{b^2}+\frac{1}{2}m^2 \frac{A_z^2}{b^2} \right) + 2\omega m^2,
\label{rhoa1} 
\end{eqnarray}
\begin{eqnarray}
\frac{\ddot{a}}{a}+\frac{\ddot{b}}{b}+\frac{\dot{a}}{a}\frac{\dot{b}}{b}=
8\pi G \left(-\frac{1}{2}\frac{\dot{A_z}^2}{b^2}+\frac{1}{2}m^2\frac{A_z^2}{b^2}\right)+ 2\omega m^2,
\label{pressurex1}
\end{eqnarray}
\begin{eqnarray}
2\frac{\ddot{a}}{a}+\frac{\dot{a}^2}{a^2}=
8\pi G \left(\frac{1}{2}\frac{\dot{A_z}^2}{b^2}-\frac{1}{2}m^2\frac{A_z^2}{b^2}\right)+ 2\omega m^2,
\label{pressurez1}
\end{eqnarray}
where $8\pi G=\frac{64\omega^3 m^4}{ \kappa^2}$. One may note that it is not necessary to write here the reduced vector field equation \eqref{vector} once again, since it would always be satisfied by these three equations \eqref{rhoa1}-\eqref{pressurez1} due to the twice-contracted second Bianchi identity for the Einstein tensor is null $G^{\mu\nu}_{;\nu}=0$ implying $T^{\mu\nu}_{;\nu}=0$. Because we write $\kappa=\sqrt{\frac{64\omega^3 m^4}{8\pi G}}$ then the scalar field becomes $B=\sqrt{\frac{4\omega}{8\pi G}}$. It can be easily calculated from \eqref{rhoa1}-\eqref{pressurez1} that the trace of the Einstein field equations gives
 \begin{eqnarray}
R=-8\pi G \frac{m^2 A_z^2}{b^2}+8\omega m^2.
\label{trace} 
\end{eqnarray}
If $A_z$ is non-zero, then \eqref{Rdef} and \eqref{trace} are identically satisfied for $m=0$, which also implies $R=0$. One may note that in this case $m=0$ equations \eqref{rhoa1}-\eqref{pressurez1} reduce to nothing new but the Einstein-Maxwell theory, 
 \bea
 {S}=\int d^4 x \sqrt{-g} \left(-\frac{1}{16\pi G}R-\frac14
F_{\mu\nu} F^{\mu\nu}\right),
\label{grlike}
\eea
for LRS Bianchi type I spacetime. Hence in this case, in contrast to what we found in the previous section where we considered the time-like component of the vector field, we are not able to obtain an interesting cosmological solution.

However, it is still possible to look for interesting cosmological dynamics once we choose a spatially homogeneous but not necessarily anisotropic spacetime metric that allows energy flux density/heat flow. To make it clear, let us now further discuss the constraints on the corresponding stress tensor of the Stueckelberg fields that arises due to our metric choice. The off diagonal components of the energy-momentum tensor, energy flux density/heat flow, of the Stueckelberg fields are obtained, from \eqref{einsteintensor}, as follows: 
 \begin{eqnarray}
 \label{flux}
T_{0i} = \frac{2A_i}{4\omega m^2}\left(f\nabla_0 R+R\nabla_0 f\right) +2\nabla_0 f A_i+(\partial _0 B-mA_0)(-mA_i),
 \end{eqnarray}
where $i$ runs for spatial coordinates, namely, $i=1,2,3$. The equation for the scalar potential ($\mu=0$) can be given from \eqref{vectoreq} as follows:
\begin{eqnarray}
\label{Azero}
\frac{1}{4\omega m^2}\left(f\nabla^0 R+R\nabla^0 f\right) +\nabla^0(\nabla_{\alpha}A^{\alpha})+m^2A^0=0.
 \end{eqnarray}
Next substituting \eqref{Azero} into \eqref{flux}, we obtain the following useful relation:
 \begin{eqnarray}
 \label{together}
T_{0i} =m \left(\partial_0B-m A_0\right) A_i.
 \end{eqnarray} 
Because we considered LRS Bianchi type I metric, then it was inevitable that $T_{0i}=0$ since the off-diagonal components of the Einstein tensor of this metric are null, i.e., $G_{0i}=0$, implying that because $G_{\mu \nu} \propto T_{\mu\nu}$ the symmetry properties of this metric do not allow a consideration of energy flux/heat flow. Because $T_{0i}=0$ in this case, equation \eqref{together} reads $A_i=0$ or $m=0$, or $mA_0=\dot{B}$. It might be interesting to note in the latter case that it is the condition $mA_0=\dot{B}$ we found in the RW spacetime solution we gave in subsection \ref{isotropic}, but now $A_i$ is not necessarily null. In other words, $A_i$ and $A_0$ can live together in this model, while it is most common in the relevant literature that either $A_i$ or $A_0$ vanishes. For instance, if we consider a potential in the form $V(A_\mu A^{\mu})$ or a non-minimal coupling to gravity in the form $R A_\mu A^{\mu}$ then they contribute to $T_{0i}$ equation as $V'(A^2)A_0 A_i$ or $R A_0A_i$ respectively. Hence if $T_{0i}$ is null, due to the chosen metric or assumed to be null by hand, then either $A_i$ or $A_0$ should be null. 

It follows from the above discussion that considering non-zero $T_{0i}$ cases, i.e., considering the presence of energy flux/heat flow, which can be implemented by using a metric that allows it, say, Bianchi type V spacetime may be interesting. In such a case equation \eqref{together} does not oblige the scalar field to be constant so the model can present rich opportunities, namely, the function $f$ again depends on the dynamics of scalar and vector fields as in the isotropic solution we gave in \ref{isotropic}. Let us consider, for instance, the simplest anisotropic generalisation of spatially open RW spacetime, i.e., LRS Bianchi type V metric,
\begin{eqnarray}
{\rm d}s^2={\rm d}t^2-e^{2nz}a^2\left({\rm d}x^2+{\rm d}y^2\right)-b^2{\rm d}z^2,
\end{eqnarray}
where $n$ is a real constant. In this case, using \eqref{eqn:a000zN} we find
\begin{eqnarray}
\label{eqn:fff}
f=mB-2n \frac{A_z}{b^2},
\end{eqnarray}
and it is easy to see that energy flux/heat flow along the $z$-axis is allowed, namely, $T_{03}=-m \dot B A_z$ is not necessarily null, since $G_{03}=2n(\frac{\dot{a}}{a}-\frac{\dot{b}}{b})$ for this metric. One may check that $G_{03} =\frac{4\omega m^2}{f^2}$ $T_{03}$, equation for LRS Bianchi type I metric given in \eqref{toget} reads
\begin{eqnarray}
\label{eqn}
2n \left(\frac{\dot a}{a}-\frac{\dot b}{b}\right)=\frac{4\omega m^2}{\left(mB-2n \frac{A_z}{b^2}\right)^2}\left(m\dot B A_z\right),
\end{eqnarray}
for LRS Bianchi type V metric. This equation explicitly shows that neither the scalar field nor the vector field even the photon mass need to be constant for $A_0=0$ anymore. In Bianchi V metric, the theory regains the function $f$ which includes both scalar and vector fields. While Bianchi type I metric choice reduces our model exactly to the Einstein-Maxwell theory, Bianchi type V metric choice, which allows energy flux/heat flow, seems to provide novel cosmological models relying on our non-minimally coupled Stueckelberg mechanism.

In this mini review, we are aimed at reviewing briefly the pros and cons of the Stueckelberg model in curved spacetime for isotropic and anisotropic spaces and hence giving a theoretical insight about the cosmological applications of the Stueckelberg mechanism. It is noteworthy that the choice of the components of the vector field and the metric changes properties of the cosmological model predictions drastically. We carried out a cosmological discussion considering an action in which the Stueckelberg fields are coupled to gravity in a specific non-minimal way and our findings are interesting enough to tempt us to further study such cosmological models, maybe, by considering different kinds of couplings between gravity and Stueckelberg fields, perhaps inspired by fundamental theories such as string theory. On the observational side, low frequency space based observatories expect to set competitive limits to the photon mass from astrophysical and cosmological data. Thus it seems that the possibility of photon being massive, and hence the Stueckelberg mechanism, both theoretically and observationally, and with its cosmological implications in particular, would attract increasing attention in the future.

\vskip 1cm

\section*{Acknowledgements}
We thank Dieter Van den Bleeken for his useful comments. The authors further thank to the anonymous referee as well as Alessandro D.A.M. Spallicci for the helpful and constructive comments that greatly contributed to improving the final version of the paper. \"{O}.A. acknowledges the support by the Science Academy in scheme of the Distinguished Young Scientist Award (BAGEP). \"{O}.A. acknowledges further the financial support he received from, and hospitality of the Abdus Salam International Centre for Theoretical Physics (ICTP), where parts of this work were carried out. N. K. acknowledges the support she has been receiving from Bo\u{g}azi\c{c}i University Scientific Research Fund with BAP project no: 7128 while this research was carrying out. N.K. acknowledges also the post-doctoral research support she is receiving from the {\.I}stanbul Technical University. 


\vskip 1cm

\begin{thebibliography}{}
\bibitem{Weinberg:1979pi} 
Weinberg, S.: Conceptual Foundations of the Unified Theory of Weak and Electromagnetic Interactions. Review of Modern Physics 52, 515 (1980), Science 210, 1212 (1980). 
\bibitem{Feldman:1963zz} 
 Feldman, G., Matthews, P.T.: Massive Electrodynamics. Phys. Rev. Lett. 130, 1633 (1963).
\bibitem{itzykson80} 
 Itzykson, C., Zuber, J.B.: Quantum Field Theory. New York, McGraw-Hill (1980).
\bibitem{Okun:1991nr} 
Okun, L.B.: The Problem of mass: From Galilei to Higgs, Proceedings, Physics at the highest energy and luminosity to understand the origin of mass. Moscow Inst. Theor. Exp. Phys. 1, (1991).
\bibitem{Okun:2006pn} 
Okun, L.B.: Photon: History, mass, charge. Acta Phys. Pol. B 37, 565 (2006). arXiv:hep-ph/0602036
\bibitem{Einstein:1905cc} 
 Einstein, A.: \" Uber einen die Erzeugung und Verwandlung des Lichtes betreffenden heuristischen Gesichtspunkt. Ann. Phys. 322, 132 (1905).
\bibitem{Einstein:1917zz} 
Einstein, A.: Zur Quantentheorie der Strahlung, Physikalische Zeitschrzft 18, 121 (1917).
\bibitem{deBroglie:1922zz} 
de Broglie, L.: Rayonnement noir et quanta de lumi\`ere. J. Phys. Radium 3, 422 (1922). 
\bibitem{debroglie23} 
de Broglie, L.: Ondes et quanta. C. R. Acad. Sci. Paris 177, 507 (1923).
 \bibitem{debroglie34} de Broglie, L.: Nouvelles recherches sur la lumi\`ere. Hermann, Paris, (1936).
 \bibitem{debroglie40} de Broglie, L.: La m\'ecanique ondulatoire du photon, Une nouvelle th\' eorie de la lumi\` ere. Hermann, Paris, (1940).
\bibitem{DeBroglie:1972hj} 
de Broglie, L., Vigier, J.P.: Photon Mass and New Experimental Results on Longitudinal Displacements of Laser Beams near Total Reflection. Phys. Rev. Lett. 28, 1001 (1972).
 \bibitem{schro55} Bass, L., Schr\" odinger, E.: Must the Photon Mass be Zero? Proc. R. Soc. A 232, 1 (1955).
\bibitem{debroglie-thesis} de Broglie, L.: Recherches sur la th\'eorie des quantas. Universit\'e de Paris-Sorbonne, Paris, (1924).
\bibitem{Proca:1900nv} Proca, A.: Sur la th\'eorie ondulatoire des \'electrons positifs et n\'egatifs. J. Phys. et Radium 7, 347, 15 (1936). 
\bibitem{proca36a} Proca, A.: Sur les photons et les particules charge pure. C. R. Acad. Sci. Paris 203, 709 (1936). 
\bibitem{proca36b} Proca, A.: Sur la th\'eorie du positon. C. R. Acad. Sci. Paris 202, 1366 (1936). 
\bibitem{proca36d} Proca, A.: Sur les \'equations fondamentales des particules \'elementaires, C. R. Acad. Sci., Paris 202, 1490 (1936). 
\bibitem{proca37} Proca, A.: Particules libres photons et particules charge pure. J. Phys. Radium 8, 23 (1937).
\bibitem{proca88} Proca, A., Proca, G.A.: Alexandre Proca 1897-1955: Oeuvre scientifique publi\'ee. (Editions Georges A. Proca,
Paris, 1988).
\bibitem{borne01} Borne, T., Lochak, G., Stumpf, H.: Quantum Field Theory and the Structure of Matter. Springer Science Business Media, (2001).
\bibitem{stueckelberg:1957} 
Stueckelberg, E.C.G.: Th\'eorie de la radiation de photons de masse arbitrairement petite. Helv. Phys. Acta 30, 209 (1957).
\bibitem{Ruegg:2003ps}
 Ruegg, H., Altaba, M.R.: The Stueckelberg field. Int. J. Mod. Phys. A 19, 3295 (2004). arXiv:hep-th/0304245
\bibitem{Lowenstein:1972pr} 
Lowenstein, J. H., Schroer, B.: Gauge invariance and Ward identities in a massive vector meson model. Phys. Rev. D 6, 54 (1972).
\bibitem{vanHees:2003dk} 
van Hees, H.: The renormalizability for massive abelian gauge field theories re-visited. (2003). arXiv:hep-th/0305076.
\bibitem{Kibble65} Kibble, T.W.: Broken symmetries. Proceedings, International Conference on High Energy Physics. Oxford University Press, (1965).
\bibitem{taylor76} Taylor, J.C.: Gauge theories of weak interactions. Cambridge University Press, Cambridge (1976).
\bibitem{Pauli:1941zz} Pauli, W.: Relativistic field theories of elementary particles. Rev. Mod. Phys. 13, 203 (1941).
\bibitem{Delbourgo:1975uf} 
Delbourgo, R.: A supersymmetric Stueckelberg formalism. J. Phys. G 8, 800 (1975).
\bibitem{Guerdane:1991at} Guerdane, M., Lagraa, M.: Canonical quantization of the massive vector supermultiplet-an example of higher-order derivative model. Zeitschrift f\" ur Physik 51, 675 (1991).
\bibitem{Deser:1974cz} 
 Deser, S., Van Nieuwenhuizen, P.: One-loop divergences of quantized Einstein-Maxwell fields. Phys. Rev. D 10, 401 (1974).
\bibitem{Stelle:1976gc} Stelle, K.S.: Renormalization of higher-derivative quantum gravity. Phys. Rev. D 16, 953 (1977).
\bibitem{Sezgin:1981xs} Sezgin, E., Van Nieuwenhuizen, P.: New ghost-free gravity Lagrangians with propagating torsion. Phys. Rev. D 21, 3269 (1979). 
\bibitem{Kallosh:1990vq} 
Bergshoeff, E., Kallosh, R.: BRST quantization of the Green-Schwarz superstring. Nucl. Phys. B 333, 605 (1990).
\bibitem{Fisch:1989rn} 
Fisch, J.M.L., Henneaux, M.: A note on the covariant BRST quantization of the superparticle. ULB-TH2/89-04-REV (1989).
\bibitem{Bergshoeff:1990vg} Bergshoeff, E., Kallosh, R.: Unconstrained BRST for superparticles. Phys. Lett. B 240, 105 (1990).
\bibitem{Delbourgo:1975aj} Delbourgo, R., Salam, A.: The Stueckelberg formalism for spin two. Nuovo Cimento 12, 297 (1975).
\bibitem{ArkaniHamed:2002sp} Arkani-Hamed, N., Georgi, H., Schwartz, M.D.: Effective field theory for massive gravitons and gravity in theory space, Ann. Phys. 305, 96 (2003). arXiv:hep-th/0210184
\bibitem{Kalb:1974yc} Kalb, M., Ramond, P.: Classical direct interstring action. Phys. Rev. D 9, 2273 (1974). 
\bibitem{Marshall:1974wf} Marshall, C., Ramond, P.: Field theory of the interacting string: The closed string. Nucl. Phys. B 85, 375 (1975).
\bibitem{vanDam:1970vg} Van Dam H., Veltman, M.: Massive and massless Yang-Mills and gravitational fields, Nucl. Phys. B 22, 397 (1970).
\bibitem{Zakharov:1970cc} Zakharov, V.I.: Linearized gravitation theory and the graviton mass. JETP Lett.12, 312 (1970).
\bibitem{Boulware:1973my} Boulware, D.G., Deser, S.: Can Gravitation Have a Finite Range?, Phys. Rev. D 6, 3368 (1972).
\bibitem{Kogan:2000uy} Kogan, I.I., Mouslopoulos, S., Papazoglou, A.: The $m \rightarrow 0$ limit for massive graviton in $dS_4$ and $AdS_4$ How to circumvent the van Dam-Veltman-Zakharov discontinuity. Phys. Lett. B 503, 173 (2001). arXiv:hep-th/0011138
\bibitem{Porrati:2000cp} 
Porrati, M.: No Van Dam-Veltman-Zakharov Discontinuity in Ads Space. Phys. Lett. B 498, 92 (2001). arXiv:hep-th/0011152
\bibitem{Hinterbichler:2011tt} Hinterbichler, K.: Theoretical aspects of massive gravity. Rev. Mod. Phys. 84, 671 (2012). arXiv:1105.3735 [hep-th]
\bibitem{Belokogne:2015etf} Belokogne, A., Folacci, A.: Stueckelberg massive electromagnetism in curved spacetime: Hadamard renormalization of the stress-energy tensor and the Casimir effect, Phys. Rev. D 93, 044063 (2016). arXiv:1512.06326 [gr-qc]


\bibitem{Glavan:2013mra} Glavan, D., Prokopec, T., Prymidis, V.: Backreaction of a massless minimally coupled scalar field from inflationary quantum fluctuations. Phys. Rev. D 89, 2, 024024 (2014). arXiv:1308.5954 [gr-qc]

\bibitem{Glavan:2015cut} Glavan, D., Prokopec T., Takahashi T.: Late-time quantum backreaction of a very light non-minimally coupled scalar. Phys. Rev. D 94, 084053 (2016). arXiv:1512.05329 [gr-qc]

\bibitem{Prokopec:2002jn} Prokopec, T., T\" ornkvist, O., Woodard, R.P.: Photon mass from inflation. Phys. Rev. Lett. 89, 101301 (2002). arXiv:astro-ph/0205331
\bibitem{Prokopec:2002uw} Prokopec, T., T\" ornkvist, O., Woodard R.P.: One loop vacuum polarization in a locally de Sitter background. Ann. Phys. 303, 251 (2003). arXiv:gr-qc/0205130
\bibitem{Prokopec:2003bx} Prokopec, T., Woodard R.P.: Vacuum polarization and photon mass in inflation. Am. J. Phys. 72 60 (2004). arXiv:astro-ph/0303358
\bibitem{Chimento:1990dk} Chimento, L. P., Cossarini, A. E.: Energy-momentum tensor renormalization for vector fields in Robertson-Walker backgrounds. Phys. Rev. D 41, 3101 (1990).
\bibitem{Frob:2013qsa} Fr\" ob, M.B., Higuchi, A.: Mode-sum construction of the two-point functions for the Stueckelberg vector fields in the Poincare patch of de Sitter space. J. Math. Phys. 55, 062301 (2014). arXiv:1305.3421[gr-qc]
\bibitem{Akarsu:2014eaa} Akarsu, \" O., Ar{\i}k, M., Kat{\i}rc{\i}, N., et al.: Accelerated expansion of the Universe \` a la the Stueckelberg mechanism. J. Cosmol. Astropart. Phys. 1407, 009 (2014). arXiv:1404.0892 [gr-qc]
\bibitem{Kouwn:2015cdw} Kouwn, S., Oh, P., Park, C.G.: Massive Photon and Dark Energy. Phys. Rev. D 93, 083012 (2016). arXiv:1512.00541[gr-qc]
\bibitem{Belokogne:2016dvd} Belokogne, A., Folacci, A., Queva, J.: Stueckelberg massive electromagnetism in de Sitter and anti-de Sitter spacetimes: Two-point functions and renormalized stress-energy tensors. (2016). arXiv:1610.00244[gr-qc]
\bibitem{Kostelecky:1991ak} Kostelecky V.A., Potting R.: CPT and Strings, Nucl. Phys. B 359 545 (1991). 
\bibitem{Kostelecky:1990pe} Kostelecky V.A., Samuel S.: Photon and Graviton Masses in String Theories, Phys. Rev. Lett. 66 1811 (1991).
\bibitem{Goldhaber:2008xy}  Goldhaber A.S., Nieto, M.M.: Photon and Graviton Mass Limits. Rev. Mod. Phys. 82, 939 (2010). arXiv:0809.1003 [hep-ph]
\bibitem{Williams:1971ms} Williams, E.R., Faller, J.E., Hill H.A.: New Experimental Test of Coulomb's Law: A Laboratory Upper Limit on the Photon Rest Mass, Phys. Rev. Lett. 26, 721 (1971). 
\bibitem{Fischbach:1994ir} Fischbach, E., Kloor, H., Langel, R.A., et al.: New geomagnetic limits on the photon mass and on long-range forces coexisting with electromagnetism. Phys. Rev. Lett. 73, 514 (1994).
\bibitem{Davis:1975mn} Davis Jr. L., Goldhaber, A.S., Nieto, M.M.: Limit on the Photon Mass Deduced from Pioneer-10 Observations of Jupiter's Magnetic Field, Phys. Rev. Lett. 35,1402 (1975).
\bibitem{Ryutov:1997zz} Ryutov, D.D.: The Role of Finite Photon Mass in Magnetohydrodynamics of Space Plasmas. Plasma Phys. Control. Fusion 39, A 73 (1997).
\bibitem{Eidelman:2004wy} Eidelman, S., et al. [Particle Data Group Collaboration]: Review of particle physics. Particle Data Group. Phys. Lett. B 592, 1 (2004).
\bibitem{Ryutov:2007zz} 
Ryutov, D.D.: Using Plasma Physics to Weigh the Photon. Plasma Phys. Control. Fusion 49, B 429 (2007).
\bibitem{Amsler:2008zzb} 
  Amsler, C., et al. [Particle Data Group Collaboration]: Review of Particle Physics. Phys. Lett. B 667, 1 (2008).
\bibitem{Olive:2016xmw} 
Patrignani, C., et al. [Particle Data Group Collaboration]: Review of Particle Physics. Chin. Phys. C 40, 100001 (2016).
\bibitem{Retino:2013gga} Retin\`o, A., Spallicci A.D.A.M., Vaivads, A.: Solar wind test of the de Broglie-Proca massive photon with Cluster multi-spacecraft data, Astropart. Phys. 82, 49 (2016). arXiv:1302.6168 [hep-ph]
\bibitem{Bonetti:2016cpo} Bonetti, L., Ellis, J., Mavromatos N.E., et al.: Photon mass limits from Fast Radio Bursts. Phys. Lett. B 757, 548 (2016). arXiv:1602.09135 [astro-ph.HE] 
\bibitem{Chibisov:1976mm}
 Chibisov, G.V.: Astrophysical Upper Limits on the Photon Rest Mass. Phys. Usp. 19, 624 (1976). 
\bibitem{Lakes:1998mi} 
 Lakes, R.: Experimental Limits on the Photon Mass and Cosmic Magnetic Vector Potential. Phys. Rev. Lett., 80, 1826 (1998). 
\bibitem{Adelberger:2003qx} Adelberger, E., Dvali, G., Gruzinov, A.: Photon-Mass Bound Destroyed by Vortices. Phys. Rev. Lett. 98, 010402 (2007). arXiv:hep-ph/0306245
\bibitem{Goldhaber:1971mr} Goldhaber A.S., Nieto, M.M.: Terrestrial and Extraterrestrial Limits on The Photon Mass. Rev. Mod. Phys. 43, 277 (1971). 
\bibitem{Wu:2016brq} 
Wu X.F., Zhang S.B., Gao, H.: Constraints on the Photon Mass with Fast Radio Bursts. Astrophys. J. 822, L 15 (2016). arXiv:1602.07835 [astro-ph.HE]
\bibitem{Bentum:2016ekl} 
Bentum, M., Bonetti, L., Spallicci, A.D.A.M.: Dispersion by pulsars, magnetars and massive electromagnetism at very low radio frequencies. to appear in Adv. Space. Res., doi:10.1016/j.asr.2016.10.018. arXiv:1607.08820 [astro-ph.IM]
\bibitem{Lowenthal:1973ka} Lowenthal, D.D.: Limits on the photon mass. Phys. Rev. D 8, 2349 (1973).
\bibitem{Tu:2005ge} Tu, L. C., Luo, J., Gillies, G. T.: The mass of the photon, Rep. Prog. Phys. 68, 77 (2005).
\bibitem{Spavieri:2011zz} Spavieri, G., Quintero, J., Gillies, G.T., et al.: A survey of existing and proposed classical and quantum approaches to the photon mass. Eur. Phys. J. D 61, 531 (2011).
\bibitem{Heeck:2013cfa} Heeck, J.: How stable is the photon? Phys. Rev. Lett. 111, 021801 (2013). arXiv:1304.2821 [hep-ph]
\bibitem{Carroll:1989vb} 
Carroll, S.M., Field, G.B., Jackiw, R.: Limits on a Lorentz and Parity Violating Modification of Electrodynamics. Phys. Rev. D 41, 1231 (1990).
\bibitem{Bonetti:2016vrq} Bonetti, L., dos Santos Filho, L.R., Helay\" el-Neto, J.A., et al.: Effective photon mass from Super and Lorentz symmetry breaking. Phys. Lett. B 764, 203 (2017). arXiv:1607.08786 [hep-ph]
\bibitem{Pani:2012vp} Pani, P., Cardoso, V., Gualtieri, L., et al.: Black-Hole Bombs and Photon-Mass Bounds. Phys. Rev. Lett. 109, 131102 (2012). arXiv:1209.0465 [gr-qc]
\bibitem{Starobinsky:1980te} Starobinsky, A. A.: A New Type of Isotropic Cosmological Models Without Singularity. Phys. Lett. B 91, 99 (1980).
\bibitem{Guth:1980zm} Guth, A.H.: Inflationary universe: A possible solution to the horizon and flatness problems. Phys. Rev. D 23, 347 (1981).
\bibitem{Albrecht:1982wi} 
Albrecht A., Steinhardt, P.J.: Cosmology for Grand Unified Theories with Radiatively Induced Symmetry Breaking. Phys. Rev. Lett. 48, 1220 (1982).
\bibitem{Linde:1981mu} Linde, A.D.: A New Inflationary Universe Scenario: A Possible Solution of the Horizon, Flatness, Homogeneity, Isotropy and Primordial Monopole Problems. Phys. Lett. B 108, 389 (1982).
\bibitem{Linde:2007fr}
Linde, A.D.: Inflationary Cosmology. Lect. Notes Phys. 738, 1 (2008). arXiv:0705.0164 [hep-th]
\bibitem{Linde:2014nna} 
Linde, A.D.: Inflationary Cosmology after Planck 2013, (2014). arXiv:1402.0526 [hep-th] 
\bibitem{Martin:2013tda} 
Martin, J., Ringeval, C., Vennin, V.: Encyclop\ae dia inflationaris. Phys. Dark Univ. 5, 75 (2014). arXiv:1303.3787 [astro-ph.CO]
\bibitem{Quevedo:2002xw} 
Quevedo, F.: Lectures on string/brane cosmology. Class. Quantum Grav. 19, 5721 (2002). arXiv:hep-th/0210292
\bibitem{Riess:1998cb} 
Riess, A.G., et al. (Supernova Search Team Collaboration): Observational Evidence from Supernovae for an Accelerating Universe and a Cosmological Constant. Astron. J. 116, 1009 (1998). arXiv:astro-ph/9805201
\bibitem{Percival:2009xn} 
Percival, W.J., et al. [SDSS Collaboration]: Baryon Acoustic Oscillations in the Sloan Digital Sky Survey Data Release 7 Galaxy Sample. Mon. Not. R. Astron. Soc. 401, 2148 (2010). arXiv:0907.1660 [astro-ph.CO]
\bibitem{Bennett:2012zja} 
Bennett, C.L., et al. [WMAP Collaboration]: Nine-Year Wilkinson Microwave Anisotropy Probe (WMAP) Observations: Final Maps and Results. The Astrophysical Journal Supplement Series 208, 20 (2013). arXiv:1212.5225 [astro-ph.CO]
\bibitem{Ade:2013zuv} 
Ade P.A.R., et al. [Planck Collaboration]: Planck 2013 results. XVI. Cosmological parameters. Astron. Astrophys. A 16, 571 (2014), arXiv:1303.5076 [astro-ph.CO], (2013).
\bibitem{Zel'dovich:1968zz} Zeldovich, Y.B.: The Cosmological constant and the theory of elementary particles. Phys. Usp. 11, 381 (1968). 
\bibitem{Weinberg:1988cp} Weinberg, S.: The cosmological constant problem. Rev. Mod. Phys. 61, 1 (1989).
\bibitem{Sahni:1999gb} Sahni, V., Starobinsky, A.A.: The case for a positive cosmological $\Lambda$ term. Int. J. Mod. Phys. D 9, 373 (2000). arXiv:astro-ph/9904398
\bibitem{Peebles:2002gy} 
 Peebles, P.J.E, Ratra, B.: The Cosmological Constant and Dark Energy, Rev. Mod. Phys. 75, 559 (2003). arXiv:astro-ph/0207347
\bibitem{Copeland:2006wr} 
Copeland, E.J., Sami, M., Tsujikawa, S.: Dynamics of dark energy. Int. J. Mod. Phys. D 15, 1753 (2006). arXiv:hep-th/0603057
\bibitem{Bamba:2012cp} 
Bamba, K., Capozziello, S., Nojiri, S., et al.: Dark energy cosmology: the equivalent description via different theoretical models and cosmography tests. Astrophys. Space Sci. 342, 155 (2012). arXiv:1205.3421 [gr-qc]
\bibitem{Sahni:2014ooa} 
Sahni, V., Shafieloo, A., Starobinsky A. A.: Model independent evidence for dark energy evolution from Baryon Acoustic Oscillations. Astrophys .J. 793 L 40 (2014). arXiv:1406.2209 [astro-ph.CO]
\bibitem{Aubourg:2014yra}
Aubourg, E., et al. [BOSS Collaboration]: Cosmological implications of baryon acoustic oscillation (BAO) measurements. Phys. Rev. D 92, 123516 (2015). arXiv:1411.1074 [astro-ph.CO]
\bibitem{Delubac:2014aqe} Delubac T., et al. (BOSS Collaboration): Baryon acoustic oscillations in the Ly$\alpha$ forest of BOSS DR11 quasars. Astron. Astrophys. 59, 574 (2015). arXiv:1404.1801 [astro-ph.CO]
\bibitem{Bezrukov:2008ej} 
Bezrukov, F. L., Magnin, A., Shaposhnikov, M.: Standard Model Higgs boson mass from inflation. Phys. Lett. B 675, 88 (2009). arXiv:0812.4950 [hep-ph]
\bibitem{Aad:2012tfa} 
 Aad, G., et al. [ATLAS Collaboration]: Observation of a new particle in the search for the Standard Model Higgs boson with the ATLAS detector at the LHC. Phys. Lett. B 716, 1 (2012). arXiv:1207.7214 [hep-ex]
\bibitem{Chatrchyan:2012xdj} 
 Chatrchyan, S., et al. [CMS Collaboration]: Observation of a new boson at a mass of $125$ GeV with the CMS experiment at the LHC. Phys. Lett. B 716, 30 (2012). arXiv:1207.7235 [hep-ex]
\bibitem{Nojiri:2010wj} Nojiri, S., Odintsov, S.D.: Unified cosmic history in modified gravity: from F(R) theory to Lorentz non-invariant models. Phys. Rep. 505, 59 (2011). arXiv:1011.0544 [gr-qc]
\bibitem{Capozziello:2011et} 
Capozziello, S., De Laurentis, M.: Extended Theories of Gravity. Phys. Rep. 509, 167 (2011). arXiv:1108.6266 [gr-qc]
\bibitem{Clifton:2011jh} 
Clifton, T., Ferreira, P.G., Padilla, A., et al.: Modified Gravity and Cosmology. Phys. Rep. 513, 1 (2012). arXiv:1106.2476 [astro-ph.CO]
\bibitem{Brans:1961sx} Brans C., Dicke, R.H.: Mach's Principle and a Relativistic Theory of Gravitation. Phys. Rev. 124, 925 (1961). 
\bibitem{Hamada:2014iga} Hamada, Y., Kawai, H., Oda, K., et al.: Higgs inflation still alive. Rev. Phys. Lett. 112, 241 (2014). arXiv:1403.5043 [hep-ph]
\bibitem{Bezrukov:2014bra} Bezrukov, F., Shaposhnikov, M.: Higgs inflation at the critical point. Phys. Lett. B 734, 249 (2014). arXiv:1403.6078 [hep-ph]
\bibitem{Ford:1989me} 
Ford, L.H.:Inflation driven by a vector field. Phys. Rev. D 40 967 (1989). 
\bibitem{Koivisto:2005mm} 
Koivisto, T., Mota, D.F.: Dark energy anisotropic stress and large scale structure. Phys. Rev. D 73, 083502 (2006). arXiv:astro-ph/0512135
\bibitem{Dimopoulos:2007ns} 
Dimopoulos, K.: Density Perturbations in the Universe from Massive Vector Fields. AIP Conf. Proc. 957, 387 (2007). arXiv:0709.1109 [hep-th] \bibitem{Bamba:2008ja} 
Bamba, K., Odintsov, S.D.: Inflation and late-time cosmic acceleration
in non-minimal Maxwell-F(R) gravity and the generation of large-scale magnetic fields. J. Cosmol. Astropart. Phys. 0804, 024 (2008). arXiv:0801.0954 [astro-ph]
\bibitem{Jimenez:2008au} 
Jim\'{e}nez J. B., Maroto A.L.: A cosmic vector for dark energy, Phys. Rev. D 78, 063005 (2008). arXiv:0801.1486 [astro-ph]
\bibitem{Golovnev:2008cf} 
Golovnev, A., Mukhanov, V., Vanchurin, V.: Vector Inflation. J. Cosmol. Astropart. Phys. 0806, 009 (2008). arXiv:0802.2068 [astro-ph]
\bibitem{Koivisto:2008xf} 
Koivisto, T., Mota, D.F.: Vector field models of inflation and dark energy. J. Cosmol. Astropart. Phys. 0808, 021 (2008). arXiv:0805.4229 [astro-ph]
\bibitem{Kanno:2008gn} 
Kanno, S., Kimura, M., Soda, J., et al.: Anisotropic inflation from vector impurity. J. Cosmol. Astropart. Phys. 0808, 034 (2008). arXiv:0806.2422 [hep-ph]
\bibitem{Jimenez:2008nm} 
 Jim\'{e}nez, J.B., Maroto, A.L.: Cosmological electromagnetic fields and dark energy. J. Cosmol. Astropart. Phys. 0903, 016 (2009). arXiv:0811.0566 [astro-ph]
\bibitem{Watanabe:2009ct} 
Watanabe, M.a., Kanno, S., Soda, J.: Inflationary Universe with Anisotropic Hair. Phys. Rev. Lett. 102, 191302 (2009). arXiv:0902.2833 [hep-th]
\bibitem{Jimenez:2009sv} 
Jim\'{e}nez, J.B., Koivisto T.S., Maroto A.L., et al.: Perturbations in electromagnetic dark energy. J. Cosmol. Astropart. Phys. 10, 029 (2009). arXiv:0907.3648 [physics.gen-ph]
\bibitem{Kanno:2010nr} 
Kanno, S., Soda, J., Watanabe, M.a.: Anisotropic power-law inflation. J. Cosmol. Astropart. Phys. 1012, 024 (2010). arXiv:1010.5307 [hep-th]
\bibitem{Golovnev:2009rm} 
Golovnev, A.: Linear perturbations in vector inflation and stability issues. Phys. Rev. D 81, 023514 (2010). arXiv:0910.0173 [astro-ph.CO] 
\bibitem{Thorsrud:2012mu} 
Thorsrud, M., Mota D.F., Hervik, S.: Cosmology of a Scalar Field Coupled to Matter and an Isotropy-Violating Maxwell Field. J. High Energy Phys. 1210, 066 (2012). arXiv:1205.6261 [hep-th]]
\bibitem{Bartolo:2013msa} 
 Bartolo, N., Matarrese, S., Peloso, M., et al.: Anisotropy in solid inflation. J. Cosmol. Astropart. Phys.1308, 022 (2013). arXiv:1306.4160 [astro-ph.CO]
\bibitem{WMAP7an}
Bennett, C.L., et al. (WMAP collaboration): Seven-year Wilkinson Microwave Anisotropy Probe (WMAP) Observations: Are There Cosmic Microwave Background Anomalies?. The Astrophysical Journal Supplement Series 192, 17 (2011). arXiv:1001.4758 [astro-ph.CO]
\bibitem{Ade:2013nlj} 
Ade, P.A.R., et al. [Planck Collaboration]: Planck 2013 results. XXIII. Isotropy and statistics of the CMB., Astron. Astrophys. A 23, 571 (2014). arXiv:1303.5083 [astro-ph.CO]
\bibitem{Ade:2013vbw} 
Ade, P.A.R., et al. (Planck Collaboration): Planck 2013 results. XXVI. Background geometry and topology of the Universe. Astron. Astrophys., A 26, 571 (2014). arXiv:1303.5086 [astro-ph.CO]
\bibitem{Mariano:2012ia} 
 Mariano, A., Perivolaropoulos, L.: CMB maximum temperature asymmetry axis: Alignment with other cosmic asymmetries. Phys. Rev. D 87, 043511 (2013). arXiv:1211.5915 [astro-ph.CO]
\bibitem{Antoniou:2010gw} 
Antoniou, I., Perivolaropoulos, L.: Searching for a cosmological preferred axis: Union2 data analysis and comparison with other probes. J. Cosmol. Astropart. Phys. 12, 012 (2010). arXiv:1007.4347 [astro-ph.CO]
\bibitem{Cai:2011xs} Cai, R.G., Tuo, Z.L.: Direction dependence of the deceleration parameter. J. Cosmol. Astropart. Phys. 1202, 004 (2012). arXiv:1109.0941 [astro-ph.CO]
\bibitem{Zhao:2013yaa} 
Zhao, W., Wu, P. X., Zhang, Y.: Anisotropy of Cosmic Acceleration. Int. J. Mod. Phys. D 22, 1350060 (2013). arXiv:1305.2701 [astro-ph.CO]
\bibitem{Campanelli:2007qn}
Campanelli, L., Cea, P., Tedesco, L.: Cosmic Microwave Background Quadrupole and Ellipsoidal Universe. Phys. Rev. D 76, 063007 (2007). arXiv:0706.3802 [astro-ph]
\bibitem{Wald:1983ky} 
 Wald, R.M.: Asymptotic behavior of homogeneous cosmological models in the presence of a positive cosmological constant. Phys. Rev. D 28, 2118 (1983).
\bibitem{Moss:1986ud} 
Moss, I., Sahni, V. : Anisotropy in the Chaotic Inflationary Universe. Phys. Lett. B 178, 159 (1986).
\bibitem{Kitada:1991ih} 
Kitada, Y., Maeda, K.i.: Cosmic no hair theorem in power law inflation. Phys. Rev. D 45, 1416 (1992).
\bibitem{Barrow97}
Barrow, J.D.: Cosmological limits on slightly skew stresses. Phys. Rev. D 55, 7451 (1997). 
\bibitem{Koivisto:2007bp} 
Koivisto, T., Mota, D.F.: Accelerating cosmologies with an anisotropic equation of state. Astrophys. J. 679, 1 (2008). arXiv:0707.0279 [astro-ph]
\bibitem{Rodrigues:2007ny} 
Rodrigues, D.C.: Anisotropic cosmological constant and the CMB quadrupole anomaly. Phys. Rev. D 77, 023534 (2008). arXiv:0708.1168 [astro-ph]
\bibitem{Battye:2009ze} 
Battye, R., Moss, A.: Anisotropic dark energy and CMB anomalies. Phys. Rev. D 80, 023531 (2009). arXiv:0905.3403 [astro-ph.CO]
\bibitem{Akarsu:2008fj} 
Akarsu, \"{O}., K{\i}l{\i}n\c{c}, C.B.: LRS Bianchi type I models with anisotropic dark energy and constant deceleration parameter. Gen. Rel. Grav. 42, 119 (2010). arXiv:0807.4867 [gr-qc]
\bibitem{Akarsu:2010zm} 
Akarsu, \"{O}., K{\i}l{\i}n\c{c}, C.B.: De-Sitter expansion with anisotropic fluid in Bianchi type-I space-time. Astrophys. Space Sci. 326, 315 (2010). arXiv:1001.0550 [gr-qc]
\bibitem{Campanelli:2011uc} 
Campanelli, L., Cea, P., Fogli, G. L., et al.: Anisotropic dark energy and ellipsoidal universe. Int. J. Mod. Phys. D 20, 1153 (2011). arXiv:1103.2658 [astro-ph.CO]
\bibitem{Akarsu:2013dva} 
Akarsu, \" O., Dereli, T., Oflaz, N.: Accelerating anisotropic cosmologies in Brans-Dicke gravity coupled to a mass-varying vector field. Class. Quantum Grav. 31, 045020 (2014). arXiv:1311.2573 [gr-qc]
\bibitem{Appleby:2009za} 
 Appleby, S., Battye, R., Moss, A.: Constraints on the anisotropy of dark energy. Phys. Rev. D 81, 081301 (2010). arXiv:0912.0397 [astro-ph.CO]
\bibitem{Appleby:2012as} 
Appleby, S.A., Linder, E.V.: Probing dark energy anisotropy. Phys. Rev. D 87, 023532
(2013). arXiv:1210.8221 [astro-ph.CO]
\bibitem{Bento:1992wy} 
Bento, M.C., Bertolami, O., Moniz, P.V., et al.: On the Cosmology of Massive Vector Fields with SO($3$) Global Symmetry. Class. Quantum Grav. 10, 285 (1993). arXiv:gr-qc/9302034
\bibitem{ArmendarizPicon:2004pm} 
Armendariz-Picon, C.: Could dark energy be vector-like? J. Cosmol. Astropart. Phys. 0407, 007 (2004). arXiv:astro-ph/0405267
\bibitem{Hosotani:1984wj} 
Hosotani, Y.: Exact solution to the Einstein-Yang-Mills equation. Phys. Lett. B 147, 44 (1984).
\bibitem{Galtsov:1991un} 
Galtsov, D.V., Volkov, M. S.: Yang Mills cosmology: Cold matter for a hot universe. Phys. Lett. B 256, 17 (1991).
\bibitem{Kiselev:2004py} 
Kiselev, V.V.: Vector field as a quintessence partner. Class. Quantum Grav. 21, 3323 (2004). arXiv:gr-qc/0402095
\bibitem{Carroll:2004ai} 
Carroll, S.M., Lim, E.A.: Lorentz-violating vector fields slow the universe down. Phys. Rev. D 70, 123525 (2004). arXiv:hep-th/0407149
\bibitem{Boehmer:2007qa} 
B\" ohmer, C.G., Harko, T.: Dark energy as a massive vector field. Eur. Phys. J. C 50, 423 (2007). arXiv:gr-qc/0701029
\bibitem{Himmetoglu:2008zp} 
Himmeto\u{g}lu, B., Contaldi, C.R., Peloso, M.: Instability of anisotropic cosmological solutions supported by vector fields. Phys. Rev. Lett. 102, 111301 (2009). arXiv:0809.2779 [astro-ph]
\bibitem{Himmetoglu:2009qi} 
Himmeto\u{g}lu, B., Contaldi, C.R., Peloso, M.: Ghost instabilities of cosmological models with vector fields non-minimally coupled to the curvature. Phys. Rev. D 80, 123530 (2009). arXiv:0909.3524 [astro-ph.CO]
\bibitem{EspositoFarese:2009aj} 
Esposito-Farese, G., Pitrou, C., Uzan, J.P.: Vector theories in cosmology. Phys. Rev. D 81, 063519 (2010). arXiv:0912.0481 [gr-qc]
\bibitem{Martin:2007ue} 
Martin, J., Yokoyama, J.: Generation of Large-Scale Magnetic Fields in Single-Field Inflation. J. Cosmol. Astropart. Phys. 0801, 025 (2008). arXiv:0711.4307 [astro-ph]
\bibitem{Hervik:2011xm} Hervik, S., Mota D.F., Thorsrud, M.: Inflation with stable anisotropic hair: Is it cosmologically viable?, J. High Energy Phys.1111, 146 (2011). arXiv:1109.3456 [gr-qc]
\bibitem{Fleury:2014qfa} 
 Fleury, P., Almeida, J.P.B., Pitrou, C., et al.: On the stability and causality of scalar-vector theories, J. Cosmol. Astropart. Phys. 11, 043 (2014). arXiv:1406.6254 [hep-th]
\bibitem{Heisenberg:2014rta} Heisenberg, L.: Generalization of the Proca Action. J. Cosmol. Astropart. Phys. 1405, 015 (2014). arXiv:1402.7026 [hep-th]
\bibitem{Jimenez2016isa} 
Jim\'{e}nez, J.B., Heisenberg, L.: Derivative self-interactions for a massive vector field. Phys. Lett. B 757, 405 (2016). arXiv:1602.03410 [hep-th]
\bibitem{DeFelice:2016uil} 
De Felice, A., Heisenberg, L., Kase, R.: Effective gravitational couplings for cosmological perturbations in generalized Proca theories. Phys. Rev. D 94, 044024 (2016), arXiv:1605.05066 [gr-qc]
\bibitem{DeFelice:2016yws} 
De Felice, A., Heisenberg, L., Kase, R.: Cosmology in generalized Proca theories. J. Cosmol. Astropart. Phys. 1606, 048 (2016). arXiv:1603.05806 [gr-qc]
\bibitem{Heisenberg:2016wtr} 
 Heisenberg, L., Kase, R., Tsujikawa, S.: Anisotropic cosmological solutions in massive vector theories. J. Cosmol. Astropart. Phys. 1611, 008 (2016), arXiv:1607.03175 [gr-qc]
\bibitem{ostro1850} 
Ostrogradsky, M.: Memoires sur les equations differentielle relatives au probleme des isoperimetres. Mem. Ac. St. Petersbourg 6, 385 (1850).
\bibitem{modifiedgr} Papantonopoulos, E.: Modifications of Einstein?s theory of gravity at large distances. Lecture Notes in Physics 892 (2015).
\bibitem{Uzan:2010pm} Uzan, J.P.: Varying Constants, Gravitation and Cosmology. Living Rev. Relat. 14, 2 (2011). arXiv:1009.5514 [astro-ph.CO]
\end{thebibliography}
\end{document}